\documentclass[11pt,siamart251104]{article}
\usepackage{graphicx}
\usepackage{epstopdf}
\usepackage{amssymb}
\usepackage{amsmath}
\usepackage{color}
\usepackage{hyperref}
\usepackage[margin=1in]{geometry}

\newcommand{\p}{\partial}
\newcommand{\og}{\omega}
\newcommand{\Og}{\Omega}
\newcommand{\fl}[2]{\frac{#1}{#2}}

\newcommand{\nn}{\nonumber}
\newcommand{\ap}{\alpha}
\newcommand{\bt}{\beta}
\newcommand{\tht}{\theta}

\newcommand{\Dt}{\Delta}

\newcommand{\be}{\begin{equation}}
\newcommand{\ee}{\end{equation}}
\newcommand{\ba}{\begin{array}}
\newcommand{\ea}{\end{array}}
\newcommand{\bea}{\begin{eqnarray}}
\newcommand{\eea}{\end{eqnarray}}
\newcommand{\beas}{\begin{eqnarray*}}
\newcommand{\eeas}{\end{eqnarray*}}
\newtheorem{theorem}{Theorem}[section]
\newtheorem{lemma}{Lemma}[section]
\newtheorem{remark}{Remark}[section]
\newcommand{\bx}{{\bf x} }
\newcommand{\gm}{\gamma}
\newcommand{\vep}{\varepsilon}

\title{Mathematical and numerical studies on ground states of trapped unitary Fermi gases}

\author{
Yongyong Cai\thanks{Laboratory of Mathematics and Complex Systems (Ministry of Education), School of Mathematical Sciences, Beijing Normal University, Beijing 100875, People’s Republic of China {\tt (yongyong.cai@bnu.edu.cn)}.}, \ 
Xinran Ruan\thanks{(Corresponding author) School of Mathematical Sciences, Capital Normal University, Beijing 100048, People’s Republic of China {\tt (xinran.ruan@cnu.edu.cn)}.} \  and
Yanzhi Zhang\thanks{Department of Mathematics and Statistics, Missouri University of Science and
Technology, Rolla,  MO 65409-0020 {\tt (zhangyanz@mst.edu)}.}
}
\begin{document}
\date{}
\maketitle

\begin{abstract}
We mathematically and numerically study the ground states of unitary Fermi gases.
Starting from the three-dimensional nonlinear Schr\"{o}dinger equation that contains a quantum pressure term and an angular momentum rotation term,  we first nondimensionalize the equation and then obtain its one-dimensional and two-dimensional  counterparts in some limit regimes of the external potentials. 
Existence and uniqueness of the ground states of the unitary Fermi gases are studied with/without the angular momentum rotation term. 
We present a regularized normalized gradient flow method to compute the ground states of trapped unitary  Fermi gases. 
Our numerical results show that the quantum pressure term has a significant effect on the ground state properties. 
Specifically, with the presence of the quantum pressure term, the vortex lattices are very different from those obtained in
conventional Bose-Einstein condensation.
\end{abstract}

{\bf Key words: }  Unitary Fermi gas;  quantum pressure; angular momentum rotation; ground states;
vortex lattices.

{\bf 2020 MSC: } 35Q55, 47J10, 47J30, 81-08, 81Q05

\section{Introduction}
\label{section1}

Numerous efforts have been devoted to the study of ultracold Fermi gases since the first experimental
realization of Bose-Einstein condensates (BEC) in dilute boson gases in 1995 \cite{Anderson1995,Davis1995,Bradley1995}.
In 1999, Fermi degeneracy was first reported in \cite{DeMarco1999}. In these ultracold dilute Fermi gases,
the two-body interactions are most important and can be characterized by a single parameter, the
$s$-wave scattering length $a_s$. If $a_s\to\infty$,  it reaches the unitary regime, where the $s$-wave scattering length
is much larger than the interspacing of the particles and the range of inter atomic potential.
In such case, the Fermi gas at unitary is dilute but
strongly interacting, and exhibits  universal properties \cite{Giorgini2008,Zwerger2012}, which has attracted much attention.
Since the first observation of the
 crossover from the weakly paired Bardeen--Cooper--Schrieffer (BCS) state to the
Bose-Einstein condensate (BEC) of molecular dimers in 2002 \cite{Hara2002}, 
the unitary Fermi gases have been extensively studied.
Experimentally,  the BCS-BEC crossover  is realized by tuning the $s$-wave scattering
 length $a_s$ from $a_s \to 0^{-}$ to $a_s \to 0^+$ using Feshbach resonance technique.  
 At resonance where the scattering length
 is divergent, the unitary regime is achieved and such systems are meta-stable
states in the BCS-BEC crossover \cite{Giorgini2008}.  
In \cite{Zwierlein2005}, quantized vortices have been observed in rotating gases of $^6$Li  atoms, which manifests the  superfluidity of unitary
Fermi gases.
Recent experimental and theoretical advances have further enriched our understanding of these strongly interacting systems. 
In \cite{PRL132243403}, evidence of the scale invariance induced by the coexistence of the spherical
symmetry and unitary interaction is observed. 
Large-scale time-dependent density functional simulations have uncovered two distinct regimes of turbulent decay in rotating fermionic superfluids \cite{PRA105013304}.
In \cite{Nature2024PG2}, a pair-fluctuation-driven pseudogap in homogeneous unitary Fermi gases of lithium-6 atoms is observed,  providing a clear answer to the critical question of whether the pseudogap can originate from strong pairing fluctuations.  
In \cite{Nature2023PG1}, experiments with photon-mediated long-range interactions have demonstrated controllable density-wave ordering in fermionic systems.  
These developments illustrate the rich many-body behavior of unitary Fermi gases and motivate 
the study of effective nonlinear Schr\"odinger models for their ground states and vortex structures.


At zero temperature,  BEC  is well described by the mean field theory, i.e., the Gross-Pitaevskii equation (GPE) \cite{Bao2013},
but such mean field theory   fails quantitatively at the BCS-BEC crossover, especially at the unitary regime \cite{Giorgini2008,Randeria2013}.
Recently, an extended Thomas-Fermi functional has been proposed to model the unitary Fermi gas \cite{Salasnich2008,Salasnich2009,Salasnich2010,Bulgac2013},
which has been successfully applied to many ground state properties and dynamical properties of Fermi gases at unitary.
Hence, a unitary Fermi superfluid
 in a rotating frame can be described  by the  nonlinear Schr\"{o}dinger equation
(NLSE) with a quantum pressure term and an angular momentum rotation term  \cite{Salasnich2008, Lundh2009}


\be\label{ONLSE}
i\hbar\fl{\p}{\p t}\psi(\bx,t) = \left[-\fl{\hbar^2}{4m}\nabla^2 + 2V(\bx) + \xi\fl{\hbar^2}{2m}
(3\pi^2)^{2/3}|\psi|^{4/3} + (1-4\lambda)\fl{\hbar^2}{4m}\fl{\nabla^2|\psi|}{|\psi|} - \Og L_z\right]\psi(\bx,t),\quad
\ee
where $\psi(\bx,t)$ denotes a complex-valued wave function, subject to the normalization constraint $\int |\psi|^2 d\bx = N$ with $N$ being the total particle number. 
Here $\bx\in{\mathbb R}^3$ represents the Cartesian coordinates and $t \geq 0$ denotes time. 
The parameter $\hbar$ is the Planck constant, $m$ is the atomic mass, and $\xi$ and $\lambda$ are real-valued parameters. 
The scalar $\Og \in{\mathbb R}$ denotes the angular velocity, 
and $L_z = -i\hbar(x\p_y - y\p_x)$ is  the $z$-component of the angular momentum
$\bx\times(-i\hbar\nabla)$. 
The function $V(\bx)$ is a real-valued external trapping potential. 
In most experiments,  a harmonic potential is used, taking the form
\bea\label{3Dpotential}
V(\bx) = \fl{m}{2}(\og_x^2 x^2 + \og_y^2 y^2 + \og_x^2 z^2), \quad \  \bx\in {\mathbb R}^3,
\eea
where $\og_x, \og_y,  \og_z >0$ are the trapping frequencies in the $x$-, $y$- and $z$-directions,
respectively.
The second last term in (\ref{ONLSE}) represents the quantum statistical pressure due to the
Pauli exclusion principle. It  was
originally introduced by von Weizs\"{a}cker to treat surface effects in nuclei and then adopted for
quantum hydrodynamics of electrons \cite{Ancona1989,Gardner1994}.
In some literature, the equation (\ref{ONLSE}) is also called the generalized
Gross--Pitaevskii equation. 

There have been other density functionals proposed in literature for unitary Fermi gases, such as the superfluid local
density functional \cite{Bulgac2007} and the  Kohn-Sham density functional \cite{Papenbrock2005}. Here, we will consider
equation (\ref{ONLSE}), which is not only physically interesting for describing unitary fermions but also mathematically interesting
for the additional quantum pressure term. Throughout the paper, we will treat $\xi$ and $\lambda$ in (\ref{ONLSE}) as two real parameters. 
We note that $\xi$ and $\lambda$ are fixed physical constants in the case of unitary Fermi gases \cite{Salasnich2008,Salasnich2010},
but it is also of mathematical interest to consider the more general setting in which they are treated as free parameters.

Recently, there have been many studies on the properties of unitary Fermi gases based on the
NLSE (\ref{ONLSE}). In \cite{Lundh2009}, the formation of a giant
vortex in the harmonic plus quartic potential were discussed.
In \cite{Adhikari2009}, the dimension reduction of the NLSE \eqref{ONLSE} were discussed and the one-dimensional
and two-dimensional Schr\"{o}dinger equations are obtained for cigar- and disk-shaped unitary
Fermi gases, respectively. However,  the quantum pressure term was neglected, which makes it impossible to
accurately study the states with non-uniform phase, e.g. quantized vortices.

{From the computational viewpoint, simulating such nonlinear Schr\"{o}dinger-type equations is also nontrivial due to the simultaneous presence of nonlinear and quantum pressure terms. 
Among various numerical approaches, normalized gradient flow (NGF) methods have become standard tools due to their simplicity, robustness, and clear physical interpretation \cite{Bao2004,Bao2013,Chiofalo2000}.
In particular, the linearized backward Euler scheme with projection has been shown to be highly efficient and convergent for computing ground states \cite{Bao2004}.
Following the NGF framework, a series of extensions and improvements have been proposed to enhance convergence and stability under more general settings. For example, the Sobolev gradient flow methods reformulate the descent direction under Sobolev inner products to accelerate convergence \cite{Danaila2010}, gradient flow methods combined with attractive-repulsive splitting techniques have been proposed to efficiently handle models with strong higher-order interaction terms \cite{Ruan2018}, while preconditioned conjugate gradient formulations improve computational efficiency for dealing with vortice case \cite{Antoine2017,Shu2024}.
More recently, structure-preserving gradient flow approaches such as the scalar auxiliary variable method \cite{Zhuang2019} and the Lagrange-multiplier-based gradient flow method \cite{Liu2021} have been developed to ensure unconditional energy stability and accuracy for nonlinear Schr\"{o}dinger-type equations.
These developments demonstrate that the NGF methodology provides a flexible and unified framework for constructing accurate and stable numerical solvers for a wide range of nonlinear quantum models. 
However, for the unitary Fermi gas considered here, the coexistence of the quantum pressure term and the rotation term in the NLSE (\ref{ONLSE}) often leads to numerical instability when directly applying existing NGF-type methods.
In this work, we propose a regularized NGF formulation that stabilizes the computation without altering much the underlying physical structure of the model.}


The aim of the paper includes (i) to study the  existence and uniqueness of ground states of
unitary Fermi gases; (ii) to  present a numerical method for computing the ground states of
unitary Fermi gases with quantum pressure and angular momentum rotation terms; (iii) to
investigate the effects of the quantum pressure term on ground states.
It is organized as follows. In Section \ref{section2},  we first nondimensionalize the NLSE
(\ref{ONLSE}) and then obtain the effective low-dimensional NLSE in  different
potential regimes. In Section \ref{section3}, the existence and uniqueness of the ground
states are discussed for both non-rotating and rotating cases. 
Then a regularized normalized gradient flow method with Lagrange multiplier  is described in Section \ref{section4} to compute the ground states of unitary Fermi gases.
 In Section \ref{section5}, numerical results are reported  for both non-rotating and
 rotating unitary Fermi gases and some concluding remarks are made in Section \ref{section6}.

\section{Dimension reduction}
\setcounter{equation}{0}
\label{section2}

In the following, we first nondimensionalize  the NLSE (\ref{ONLSE}) and then discuss its dimension
reduction in different regimes of trapping frequencies.   
Denote the dimensionless variables $\tilde{t}$, $\tilde{\bx}$ and $\tilde{\psi}$ by
\bea
\tilde{t} = \fl{1}{\og_m}t, \qquad \tilde{\bx} = \fl{1}{a_0}\bx, \qquad \tilde{\psi}(\tilde{\bx},
 \tilde{t}) =  \sqrt{\fl{a_0^3}{N}} \, \psi(\bx, t), \qquad \tilde{\Og} = \og_m\Og,
\eea
where $\displaystyle \og_m = \max\{\og_x, \og_y, \og_z\}$ and $a_0 =\sqrt{\hbar/(2m\og_m)}$.
Substituting these relations into (\ref{ONLSE}) and, for simplicity, omitting the tildes, we obtain the following dimensionless equation: 
\bea\label{NLSE3D}
i\fl{\p\psi(\bx, t)}{\p t} = \left[-\fl{1}{2}\nabla^2  + V(\bx) + \bt|\psi(\bx,t)|^{4/3}
+\ap\fl{\nabla^2|\psi(\bx,t)|}{|\psi(\bx,t)|} - \Og L_z\right]\psi(\bx,t), \quad \bx\in{\Bbb R}^3, \ \  t\geq 0,
\eea
where $\bt =  \xi\hbar(3\pi^2N)^{2/3}/(2m)$ and $\ap = (1-4\lambda)/2$ are dimensionless parameters, 
$L_z = -i(x\p_y-y\p_x)$ is the corresponding dimensionless operator, 
and the dimensionless trapping potential is given by 
\bea\label{potential3}
V(\bx) = \fl{1}{2}\big(\gamma_x^2 x^2 + \gamma_y^2 y^2 + \gamma_z^2 z^2\big),
\quad \ \bx\in{\mathbb R}^3, 
\eea
where $\gamma_x = \og_x/\og_m$, $\gamma_y = \og_y/\og_m$ and
$\gamma_z = \og_z/\og_m$ are the dimensionless trapping frequencies. 
It is easy to verify that the wave function in (\ref{NLSE3D}) satisfies the normalization condition
\bea\label{norm}
\|\psi(\cdot, t)\|^2 = \int_{{\Bbb R}^3}|\psi(\bx,t)|^2 d\bx \equiv 1, \
\quad t\geq 0.
\eea

\noindent{\bf I. Disk-shaped condensates \ }
In a disk-shaped condensate, i.e., the condensate is strongly confined in one direction, its
properties can be effectively described by a two-dimensional (2D) NLSE \cite{Adhikari2009,Cai2013}.
Without loss of generality, we assume that the external potential
in (\ref{potential3}) is tightly confined in the $z$-direction (i.e., $\gamma_z \gg \gamma_x$  and $\gamma_z \gg \gamma_y$).

 Due to the strong confinement, we assume that the time evolution does not
include excitations in the $z$-direction. 
Thus, the wave function along the $z$-direction can be well approximated by the ground state $\phi(z)$,  which satisfies
\bea\label{eq1}
\mu_z\phi(z)=-\frac{1}{2}\p_{zz}\phi(z)+\frac{\gamma_z^2}{2}z^2\phi(z)+
\alpha\frac{\p_{zz}\left|\phi\right|}{|\phi|}\phi(z),\quad z\in{\mathbb R}, 
\eea
with $\int_{-\infty}^\infty |\phi(z)|^2dz  = 1$.
Solving (\ref{eq1}), we obtain
\beas
\phi(z)=\fl{\gamma_z^{1/4}}{(\pi)^{1/4}(1-2\alpha)^{1/8}} e^{-\frac{\gamma_zz^2}{2\sqrt{1-2\alpha}}}, \qquad
\mu_z=\frac{\gamma_z\sqrt{1-2\alpha}}{2}.
\eeas

Then we adopt the following ansatz for the full wave function 
\be\label{ansa1}
\psi(\bx,t)= e^{-i\mu_z t} \phi(z)\,\psi_2(x,y,t),\qquad \bx=(x,y,z) \in{\mathbb R}^3.
\ee
Substituting the ansatz (\ref{ansa1}) into (\ref{NLSE3D}), multiplying both sides by $\phi(z)$,  integrating with respect to $z$,
and using the normalization condition (\ref{norm}),
we obtain the corresponding {\it quasi-2D equation} for $\psi_2(x,y,t)$: 
\be\label{gpe2d}
i\fl{\p\psi_2(x, y, t)}{\p t}=\left[-\frac{1}{2}\nabla_\bot^2+V_2(x,y)+\beta_{2}|\psi_2|^{4/3}+\alpha
\frac{\nabla_\bot^2|\psi_2|}{|\psi_2|} - \Og L_z\right]\psi_2(x,y,t), \quad (x,y)\in{\mathbb R}^2, \ \ t>0, 
\ee
where $V_2(x,y)=\frac12(\gamma_x^2x^2+\gamma_y^2y^2)$, $\nabla_\bot^2 = \p_{xx}+\p_{yy}$ and 
\be\label{def:beta2}
\beta_{2} := \bt\int_{-\infty}^\infty |\phi(z)|^{10/3} dz = \beta \sqrt{\frac{3}{5}}\frac{\gamma_z^{\frac{1}{3}}} {\pi^{\frac{1}{3}}
(1-2\ap)^{\frac{1}{6}}}.
\ee

\bigskip

\noindent {\bf  II. Cigar-shaped condensate \ } When $\Og = 0$, if the potential is strongly confined in
two directions,  the equation \eqref{NLSE3D} can be reduced to a one-dimensional (1D) NLSE.
Without loss of generality, we assume that
the external potential in \eqref{potential3} is tightly confined in the $y$- and $z$-directions (i.e., $\gm_z\gg\gm_x$
and $\gm_y\gg\gm_x$).

Similarly,
due to the strong confinement, we assume that the evolution of the wave function in the $(y,z)$-
directions is well approximated by the ground state $\phi(y, z)$, which satisfies
\bea\label{eq2}
&&\mu_{y,z}\phi(y,z)=\left[-\frac{1}{2}(\p_{yy}+\p_{zz})+\frac{1}{2}(\gamma_y^2y^2+\gamma_z^2z^2)+
\alpha\frac{(\p_{yy}+\p_{zz})\left|\phi\right|}{|\phi|}\right]\phi(y,z), \quad \ (y, z)\in{\mathbb R}^2, \qquad
\eea
with $\int_{-\infty}^\infty\int_{-\infty}^\infty|\phi(y,z)|^2 dy dz = 1$. 
Solving \eqref{eq2} yields
\beas
\phi(y,z)=\fl{\gamma_y^{1/4}\gamma_z^{1/4}}{(\pi)^{1/2}(1-2\alpha)^{1/4}} e^{-\frac{\gamma_yy^2+\gamma_zz^2}{2\sqrt{1-2\alpha}}},\qquad
\mu_{y,z}=\frac{\sqrt{1-2\alpha}}{2}(\gamma_y+\gamma_z).
\eeas

Taking the ansatz
\be\label{ansa2}
\psi(\bx,t)=e^{-i\mu_{y,z} t}\phi(y,z)\psi_1(x,t), \qquad \bx = (x,y,z) \in{\mathbb R}^3,
\ee
and following a similar procedure as in the derivation of (\ref{gpe2d}), we get the following  {\it
quasi-1D equation} for $\psi_1(x,t)$:
\be\label{gpe1d}
i\fl{\p\psi_1(x,t)}{\p t}=\left[-\frac{1}{2}\p_{xx}+V_1(x)+\beta_{1}|\psi_1|^{4/3}+\alpha
\frac{\p_{xx}|\psi_1|}{|\psi_1|}\right]\psi_1(x,t), \quad  \ x\in{\mathbb R}, \quad t > 0, 
\ee
where $V_1(x)=\frac{1}{2}\gamma_x^2x^2$ and the constant
\be \label{def:beta1}
\beta_{1} := \bt \int_{-\infty}^\infty\int_{-\infty}^\infty |\phi(y,z)|^{10/3} dydz
=  \frac{3\beta\gamma_y^{1/3}\gamma_z^{1/3}}{5\pi^{2/3}(1-2\alpha)^{1/3}}.
\ee
We remark here that in \cite{Adhikari2009}, the dimension reduction is performed using different
approaches in which the quantum pressure term is omitted. 
However, our numerical results indicate that the quantum pressure term plays an important role in determining ground state properties, 
particularly when the nonlinearity coefficient $\bt_d$ is small or when  quantum vortices are present in the 
ground state. More detailed discussions are provided in Section \ref{section5}.
\bigskip

In the sequel, we rewrite (\ref{NLSE3D}), (\ref{gpe2d}) and (\ref{gpe1d}) in a unified
form and consider the general $d$-dimensional  NLSE:
\bea\label{NLSED}
i\fl{\p\psi(\bx,t)}{\p t}=\left[-\frac12\nabla^2+V(\bx)+\beta|\psi|^{4/3}+\ap\frac{\nabla^2
|\psi|}{|\psi|}-\Og L_z\right]\psi(\bx, t),\quad\  \bx\in\Bbb R^d,\quad t >0,\quad
\eea
where $d = 1, 2, 3$ when $\Og = 0$ and  $d =  2, 3$ when $\Og \neq 0$, $\bt$ is a general constant, and $V(\bx)$ is a general external potential. 
There are two important invariants associated with the NLSE \eqref{NLSED}: the mass 
\bea\label{norm}
\|\psi(\cdot, t)\|^2 = \int_{{\Bbb R}^d}|\psi(\bx,t)|^2 d\bx \equiv
 \|\psi(\cdot, 0)\|^2  = 1, \quad\ \  t\geq 0,
\eea
and the energy
\bea\label{energy}
E(\psi(\cdot,t)) &=& \int_{\Bbb R^d} \left[\fl{1}{2}|\nabla\psi|^2 + V(\bx)|\psi|^2 + \fl{3}{5}\bt|\psi|^{10/3}
-\ap\left|\nabla|\psi|\right|^2 - \Og{\rm Re}(\bar{\psi}L_z\psi)\right]d\bx\qquad\qquad \nn\\
&\equiv& E(\psi(\cdot, 0)),
\quad\ \ t\geq0,
\eea
where ${\rm Re}(f)$ and $\bar{f}$ represent the real part and the complex conjugate of a function
$f$, respectively.

\section{Ground states}
\label{section3}
\setcounter{equation}{0}

The ground state $\phi_g(\bx)$ can be obtained by minimizing the energy functional
(\ref{energy}) under the normalization constraint (\ref{norm}),
i.e.,   find $\phi_g(\bx) \in S_d$ such that
\begin{equation}\label{minimize}
    E_g := E\left(\phi_g\right) = \min_{\phi \in S_d}
    E(\phi),
\end{equation}
where the nonconvex set $S_d$ is defined as
\bea\label{nonconset}
S_d:=\left\{ \phi\in H^1(\Bbb R^d) \ | \ \|\phi\|=1,\
E(\phi)<\infty \right\},\quad \  \ d=1,2, 3.
\eea
It is easy to see that the ground state $\phi_g(\bx)$ defined in (\ref{minimize}) satisfies the
following Euler-Lagrange equation
\bea\label{Euler-Lag}
\mu\,\phi(\bx) = \left[-\fl{1}{2}\nabla^2 + V(\bx) + \bt|\phi(\bx)|^{4/3} + \ap\fl{\nabla^2|\phi(\bx)|}
{|\phi(\bx)|} - \Og L_z \right]\phi(\bx), \quad  \ \ \bx\in{\mathbb R}^d
\eea
where $\|\phi\|^2 = 1$, and the eigenvalue (also called the chemical potential) is given by 
\bea \label{defn:mu}
\mu := \mu(\phi) = \int_{\Bbb R^d} \left[\fl{1}{2}|\nabla\phi|^2 + V(\bx)|\phi|^2 +
\bt|\phi|^{10/3} -\ap\left|\nabla|\phi|\right|^2 - \Og{\rm Re}(\bar{\phi} L_z\phi)\right]d\bx.
\eea
The eigenfunctions $\phi(\bx)$ of the constrained nonlinear eigenvalue problem (\ref{Euler-Lag}) are
usually referred to as stationary states. 
Among them,  the state with the lowest energy is called the {\it ground state}, 
whereas those with higher energies are referred to as {\it excited states}.

\subsection{Existence and uniqueness of ground states}

To study existence and uniqueness of ground states of unitary Fermi gases, we first
present the following lemmas:
\begin{lemma}\label{lem1}
Let $\alpha<\frac12$ and $\Omega=0$ in (\ref{NLSED}).  
For any  function $\phi(\bx)$ satisfying
$E(\phi)<\infty$ in (\ref{energy}), there is
 \be\label{pp}
 E(|\phi|)\leq E(\phi),
 \ee
and  the equality in (\ref{pp}) holds iff $\phi(\bx)=e^{i\theta}|\phi(\bx)|$, where
$\theta\in{\mathbb R}$ is a constant.
\end{lemma}
{\it Proof.}
This follows from the inequality \cite{Lieb}
\be \|\nabla |f|\| \leq \|\nabla f\|,
\ee
where the equality holds iff $f=e^{i\theta}|f|$ for some constant $\theta\in\Bbb R$.
\hfill$\Box$

\begin{lemma}\label{lem2} Let $\beta\ge0$, $\Omega=0$ and $\alpha<\frac12$ in (\ref{NLSED}).
 For $\sqrt{\rho}\in S_d$ in (\ref{nonconset}), we have $E(\sqrt{\rho})$ is strictly convex in $\rho$.
\end{lemma}
{\it Proof.}
This immediately follows from  \cite{Lieb} by noticing that $f(s)=s^{5/3}$ ($s>0$) is convex in $s$.
\hfill$\Box$

\begin{theorem}\label{thm:gs} Let  $\Og = 0$ in (\ref{NLSED}).  When  $\alpha<\frac12$,
$V(\bx)\ge 0$ for $\bx\in{\mathbb R}^d$ and $\lim\limits_{|\bx|\to\infty}V(\bx)=\infty$,  there exists a  ground state $\phi_g(\bx)\in S_d$,  if one
of the following conditions holds:
\begin{enumerate}
\item[(i)]  $d = 1$ or $2$;
\item[(ii)] $d=3$ and  $\beta> -C_b(\frac12-\alpha)$, where the constant
\be\label{cb}
C_b=\inf_{f\neq0,\,f\in H^1(\Bbb R^3)}\frac{\|\nabla f\|^2\|f\|^{4/3}}{\|f\|_{10/3}^{10/3}}.
\ee
\end{enumerate}
Moreover,  the function $\phi(\bx)=e^{i\theta}|\phi_g(\bx)|$ is also a ground state, for any constant
 $\theta\in\Bbb R$.  That is, the positive ground state $|\phi_g(\bx)|$ is unique if $\beta\ge0$.
In contrast, if $d=3$ and $\beta< -C_b(\frac12-\alpha)$, there exists no ground state.

\end{theorem}

\noindent {\it Proof. }
(1) Existence of ground states. First, we show that $E(\cdot)$ is bounded from below in $S_d$ under the conditions in the theorem. In one dimension, using the Sobolev inequality and Cauchy inequality, for any $\phi\in S_1$ we have
\be
\|\phi\|_{10/3}^{10/3}\leq \|\phi\|_{\infty}^{4/3}\|\phi\|^2\leq C \left(\|\phi^\prime\|\|\phi\|\right)^{4/3}\leq \vep \|\phi^\prime\|^2+C_\varepsilon,\quad \forall \vep\ge0,
\ee
where $C$ is a constant and $C_\varepsilon$ depends on $\varepsilon$. 
Thus, one can easily verify that $E(\cdot)$ is bounded from below in $S_1$ whenever $\alpha<\frac12$.  
In two dimensions, using the Sobolev inequality and Young's ineuality, we have for any $\phi\in S_2$,
\be
\|\phi\|_{10/3}^{10/3}\leq \|\phi\|^{2/3}\|\phi\|_4^{8/3}\leq \left(\|\nabla\phi\|^2\|\phi\|^2\right)^{2/3}\leq \|\nabla\phi\|^{4/3}\leq \vep \|\nabla\phi\|^2
+C_\vep,\quad \forall \vep>0,
\ee
where $C_\vep$ depends on $\vep$. Hence, similar to the one dimensional case, $E(\cdot)$ is bounded from below  in $S_2$ for $\alpha<\frac12$.  
In three dimensions, using the Sobolev inequality, for any $\phi\in S_3$, we have
\be
\|\phi\|_{10/3}^{10/3}\leq \frac{1}{C_b} \|\nabla\phi\|^2 \|\phi\|^{4/3}\leq \frac{1}{C_b} \|\nabla\phi\|^2.
\ee
Therefore, whenever $\beta>-C_b(\frac{1}{2}-\alpha)$, $E(\cdot)$ is bounded from below in $S_3$. 
Consequently, the energy functional $E(\cdot)$ is bounded from below in $S_d$ for all admissible dimensions $d$.

Thus, in all cases, we can take a sequence $\left\{\phi^n\right\}_{n=1}^\infty$ minimizing the energy functional $E(\cdot)$ in $S_d$, and the sequence is uniformly bounded in $H^1$, $L^{10/3}$, and $L^2_{V}=\{\phi\,|\,\int_{\Bbb R^d}V(\bx)|\phi(\bx)|^2\,d\bx<\infty\}$. There exists $\phi^\infty\in H^1\cap L^{10/3}\cap L^2_{V}$ and a  subsequence (denoted as the original sequence for simplicity), such that
\be
\phi^n\rightharpoonup \phi^\infty,\quad \mbox{in } H^1\cap L^{10/3}\cap L^2_{V}.
\ee
The confining condition $\lim\limits_{|\bx|\to\infty}V(\bx)=\infty$ ensures that $\phi^\infty\in S_d$ \cite{Lie}. Hence $\phi^n$ converge to $\phi^\infty$ in $L^2$. By interpolation, $\phi^n\to\phi^\infty$ in $L^{10/3}$. Together with the lower semi-continuity of the $H^1$ and $L_{V}^2$ norm, we get
\be
E(\phi^\infty)\leq\liminf\limits_{n\to\infty}E(\phi^n)=\min\limits_{\phi\in S_d}E(\phi),
\ee
i.e., $\phi^\infty$ is a minimizer. From Lemma \ref{lem1}, $\phi^\infty$ can be chosen nonnegative which happens to be positive in view of the maximal principle for the associated Lagrange equation.  The uniqueness follows from the strict convexity of the energy functional $E(\cdot)$ and Lemma \ref{lem2}.

(2) Nonexistence. In  three dimensions, if $\beta< -C_b(\frac12-\alpha)$, we show there exists no ground state. Recalling that $C_b$ (\ref{cb}) can be attained at some $H^1$ function $\phi\in S_3$. Let $\phi^\vep=\vep^{-3/2}\phi(\bx/\vep)\in S_3$, then
\be
E(\phi^\vep)=\vep^{-2}(C_b(\frac12-\alpha)+\beta)\|\phi\|_{10/3}^{10/3}+\int_{\Bbb R^3}V(\vep\bx)|\phi|^2\,d\bx.
\ee
When $C_b(\frac12-\alpha)+\beta<0$, letting $\vep\to 0^+$, one finds that
$\lim\limits_{\vep\to0^+}E(\phi^\vep)=-\infty$, i.e. there exists no ground state in this case.
\hfill$\Box$

 \begin{theorem}\label{thm:gs2}  Let $\Og \neq 0$  in (\ref{NLSED}).  When $\ap < 1/2$ and
 $V(\bx)$  is a harmonic potential, {taking the form 
 \bea \label{def:pot}
\quad V(\bx)=\left\{\begin{array}{ll}
\frac{\gm_x^2 x^2}{2},  & d = 1, \\
\frac{\gm_x^2 x^2 + \gm_y^2 y^2}{2},  & d = 2, \\
\frac{\gm_x^2 x^2 + \gm_y^2 y^2 + \gm_z^2 z^2}{2},  & d = 3, 
\end{array}\right.
\eea
 }
 there exists a  ground state $\phi_g(\bx)\in S_d$,  if one of the following holds
 \begin{enumerate}
 \item[(i)]   $|\Omega| < \gm_{\min}$ when $d = 2$;
 \item [(ii)] $|\Og| < \gm_{\rm min}$ and $\beta> -C_b(\frac12-\alpha)$ when $d = 3$,
  \end{enumerate}
 where $\gm_{\min}  = \min\{\gm_x, \gm_y\}$ if $d = 2,3$ and  the constant $C_b$ is  defined in (\ref{cb}).
 Otherwise,  no ground state exists   if $|\Omega|>\gamma_{\min}$ or
 $\beta_3\leq-C_b(\frac12-\alpha)$.\end{theorem}

\noindent{\it Proof. } The existence part follows from arguments analogous to those in Theorem \ref{thm:gs}. 
Here we present the proof of the non-existence result in the 3D case. The 2D case can be treated in the same way.

 First, note that for any real-valued function $\phi\in S_3$, $\int_{\Bbb R^3}\bar{\phi} L_z\phi\,d\bx=0$. 
 Hence, if $\beta\leq-C_b(\frac12-\alpha)$, by choosing the same test functions as in Theorem \ref{thm:gs},  one easily obtains $\inf\limits_{\phi\in S_3}E(\phi)=-\infty$, i.e. no ground state exists. 
 To complete the proof for the case $d=3$,  it remains to consider the case  $|\Omega|>\min\{\gamma_x,\gamma_y, \gamma_z\}$.

Without loss of generality, we can assume that $\gamma_x\leq \gamma_y$ and $\Omega>\gamma_x$. Choose a nonnegative  $C_0^\infty(\Bbb R^2;[0,\infty))$ function $\rho(x,y)$ satisfying
 \be
 \int_{\Bbb R^2}\rho^2(x,y)dxdy=1,\quad \mbox{supp}(\rho)\subset\{(x,y)\in \Bbb R^2\big| 1\leq\sqrt{x^2+y^2}\leq 2\}
 \ee
 and
 \be
 \int_{\Bbb R^2} \frac{\gamma_x^2 x^2 + \gamma_y^2 y^2}{2} \rho^2(x,y)\,dxdy=\frac{(\gamma_x+\epsilon)^2}{2},\qquad 0<\epsilon<\frac{\Omega-\gamma_x}{2},
 \ee
 then introduce the cylindric coordinate $(r,\theta,z)$, where $r\ge0$, $\theta\in[0,2\pi)$, with $x=r\cos\theta,y=r\sin\theta$, and denote
 \be
 f_n(r,\theta,z)=e^{in\theta}\rho(x,y)w_0(z),\quad w_0(z)=\pi^{-1/4}e^{-z^2/2},\quad n\in\Bbb Z^+.
 \ee
Such $\rho$ exists as we can think $\rho^2$ as a Dirac distribution concentrated on point $(1,0)^T$ in the limiting sense. Then, using the property $\rho=0$ for $r\leq 1$, we have
\begin{align*}
&\int_{\Bbb R^3}\frac12\left(|\partial_x f_n|^2+|\partial_y f_n|^2\right)\,d\bx-\int_{\Bbb R^3}\alpha\left(|\partial_x |f_n||^2+|\partial_y |f_n||^2\right)\,d\bx\\
&=(1/2-\alpha)\pi\int_0^\infty|\p_r\rho|^2r\,dr+n^2\pi \int_0^\infty
\frac{|\rho|^2}{r^2}r\,dr\\&\leq  C_1+n^2\pi \int_0^\infty
|\rho|^2r\,dr=C_1+\frac{n^2}{2}.
\end{align*}
Furthermore, noticing the proof in Theorem \ref{thm:gs} and the property of $\rho$, we get
\begin{align*}
&-\Omega\int_{\Bbb R^3}\mbox{Re}(\bar{f_n} L_z f_n)\,d\bx=-\Omega\int_{\Bbb R^3}\mbox{Re}(\bar{f_n}(-i)\p_\theta f_n)\,d\bx
=-n\Omega,\\
&\int_{\Bbb R^3}\frac{\gamma_x^2 x^2 + \gamma_y^2 y^2}{2}|f_n|^2\,d\bx=\frac{(\gamma_x+\epsilon)^2}{2}.
\end{align*}
Set $f_n^\delta=\delta^{-1}f_n(r/\delta,\theta,z)=\delta^{-1}f_n(x/\delta,y/\delta,z)$ for $\delta>0$, then the energy $E(f_n^\delta)$  satisfies
\be
E(f_n^\delta)\leq (C_1+\frac{n^2}{2})\delta^{-2}+\delta^2 \frac{(\gamma_x+\epsilon)^2}{2}
-n\Omega+C_3\delta^{-4/3}+C_4,\quad C_1,C_3\ge0.
\ee
Choose
\be \delta_n^2=\sqrt{\frac{2C_1+n^2}{(\gamma_x+\epsilon)^2}},\ee
then
\be
\delta_n^2\ge n/\Omega, \quad \text{for sufficient large } n,
\ee
and  we have
\begin{align*}
E(f_n^{\delta_n})\leq& (\gamma_x+\epsilon)(\sqrt{2C_1+n^2})-n\Omega+C_5+C_3\Omega^{4/3}/n^{4/3}
\\\leq& (\gamma_x+\epsilon)\frac{2C_1}{\sqrt{2C_1+n^2}}-n\epsilon+C_5+C_3\Omega^{4/3}/n^{4/3}.\nn
\end{align*}
Let $n\to+\infty$, it is obvious that $E(f_n^{\delta_n})\to -\infty$, i.e. there exists no ground state.
 \hfill$\Box$ \\

 In fact, Theorem \ref{thm:gs2} can be extended to more general trapping potentials. Similar results as \cite{Seiringer2002} concerning the symmetry breaking of ground states can be obtained.
\subsection{Approximations of ground states when $\Og = 0$}
\label{section3-2}

When the external potential $V(\bx)$ is given by \eqref{def:pot} and $\Og = 0$, we have approximations of ground states in different parameter regimes.
In the strongly repulsive interacting regimes,  i.e., $\bt \gg 1$, the nonlinear term becomes dominant and thus  $\nabla^2\phi$ and $\nabla^2|\phi|$  can
be neglected in (\ref{Euler-Lag}), leading to 
\bea\label{Eq:mu}
\mu\phi(\bx) = \left[V(\bx) + \bt |\phi|^{\fl{4}{3}}\right]\phi(\bx),\quad  \bx\in{\mathbb R}^d,  
\eea
where $ \|\phi\| = 1$.
Solving (\ref{Eq:mu}) gives the Thomas--Fermi (TF) approximation of  ground states
\bea\label{TFgs}
\phi_g^{\rm TF}(\bx)  = \left\{\begin{array}{ll}
\displaystyle\left[(\mu^{\rm TF}-V(\bx))/\bt\right]^{\fl{3}{4}},   \quad &\mu^{\rm TF} \geq V(\bx), \\
0, & {\rm otherwise},
\end{array}\right.
\eea
and the corresponding chemical potentials are
\bea
\quad \mu^{\rm TF}=\left\{\begin{array}{ll}
2^{5/4}\beta^{3/4}(\gamma_x/3\pi)^{1/2},  & d = 1, \\
5^{2/5}\beta^{3/5}(\gm_x\gm_y/4\pi)^{2/5}, &d = 2, \\
2^{1/2}\beta^{1/2}(\gm_x\gm_y\gm_z/\pi^2)^{1/3}, \quad &d = 3.

\end{array}\right.
\eea

On the other hand,  in the weakly interacting regime when  $|\beta|\ll 1$,
 the nonlinear term $\bt|\phi|^{4/3}\phi$ can be neglected in (\ref{Euler-Lag}), which results in
 \bea\label{eq5}
 \mu\phi(\bx) = -\fl{1}{2}\nabla^2\phi(\bx) + V(\bx)\phi(\bx) +\ap\fl{\nabla^2|\phi|}{|\phi|}\phi(\bx),
 \quad  \bx\in{\mathbb R}^d, 
 \eea
 where $\|\phi\| = 1$.
From (\ref{eq5}), we obtain the approximations of ground states in the weakly interacting regime,
\bea\label{0gs}
\phi_{\rm ho}(\bx) = \fl{1}{\pi^{\frac{d}{4}}(1-2\ap)^{\fl{d}{8}}}\left\{\begin{array}{ll}
\gm_x^{\frac14} e^{-\frac{\gm_xx^2}{2\sqrt{1-2\alpha}}},   & d = 1, \\
(\gm_x\gm_y)^{\frac14} e^{-\fl{\gm_xx^2+\gm_yy^2}{2\sqrt{1-2\alpha}}},  & d = 2, \\
(\gm_x\gm_y\gm_z)^{\frac14} e^{-\fl{\gm_xx^2+\gm_yy^2+\gm_zz^2}{2\sqrt{1-2\alpha}}}, & d = 3, 
\end{array}\right.
\eea
and the corresponding chemical potentials are 
\bea\label{mugs}
\mu = \fl{\sqrt{1-2\alpha}}{2}\left\{\begin{array}{ll}
\gm_x,  & d = 1,\\
\gm_x + \gm_y, & d = 2, \\
\gm_x + \gm_y + \gm_z, & d = 3.
\end{array}\right.
\eea

The approximations (\ref{TFgs}) and (\ref{0gs}) can be used as initial conditions in
computing ground states, and they also serve as useful benchmarks for validating our numerical schemes. 

\section{Numerical method}
\setcounter{equation}{0}
\label{section4}

In this section, we introduce an efficient and accurate numerical method to compute ground states of unitary Fermi gases
with/without an angular momentum rotation term.  There have been numerous algorithms proposed
in literature to compute  ground states of Bose-Einstein condensation whose properties are described
by a nonlinear Schr\"{o}dinger equation without the quantum pressure term. However, to the best of our knowledge, there is few numerical method reported for finding ground states when both the quantum pressure and angular momentum rotation terms are present in NLSE, except some preliminary study \cite{Lundh2009}.
We adopt here the {\it gradient flow with discrete normalization (GFDN)} method \cite{Bao2004}, which has been shown to be robust and efficient for computing ground states of nonlinear Schrödinger-type equations.
Without loss of generality, we consider the GFDN formulation with a Lagrange multiplier \cite{Liu2021}: 
\bea\label{DNLSE}
&&\p_t \phi(\bx, t) =  \left[\fl{1}{2}\nabla^2 -V(\bx) - \bt|\phi|^{\frac{4}{3}}
 - \ap\fl{\nabla^2|\phi|}{|\phi|} + \Og L_z + \mu \right]\phi(\bx, t), \quad \bx\in{\mathbb R}^d, \ \
t\in[t_n, t_{n+1}],\qquad\quad \, \\
\label{Projection}
&&\phi(\bx, t_{n+1}) = \fl{\phi(\bx, t_{n+1}^-)}{\|\phi(\cdot, t_{n+1}^-)\|}, \quad \ \bx\in{\mathbb R}^d,
\quad\ \, \mbox{with}\quad\|\phi(\cdot, t_{n+1}^-)\|= \sqrt{\int_{{\mathbb R}^d}|\phi(\bx, t_{n+1}^-)|^2 d\bx},
\eea
where $\mu$ is the Lagrange multiplier defined in \eqref{defn:mu},  
and $t_n = n\Dt t$  for $n \in \mathbb{N}$ with $\Dt t > 0$ being the fixed time step.
Here $\phi(\bx, t_{n+1}^-)$ denotes the intermediate solution obtained from \eqref{DNLSE} at $t=t_{n+1}$. 
Equation \eqref{DNLSE} can in fact be derived by applying the steepest descent method to the energy functional \eqref{energy}. 
For more detailed discussions on GFDN, we refer readers to \cite{Bao2004}.
The initial condition for \eqref{DNLSE}--\eqref{Projection} is prescribed as 
\bea\label{DInitial}
\phi(\bx, 0) = \phi_0(\bx), \quad\ \bx\in{\mathbb R}^d, \quad \ \mbox{with \  \ $\|\phi_0(\cdot)\| = 1$.}\ \
\eea
Due to the confinement of $V_d(\bx)$, the wave function $\phi(\bx)$ decays to
zero exponentially fast  when  $|\bx|\to\infty$.  Hence, we can truncate the whole
space problem (\ref{DNLSE})--(\ref{DInitial})  into a bounded domain ${\mathcal D}\subset{\mathbb R}^d$
and consider homogeneous Dirichlet boundary conditions. In practice,  the domain ${\mathcal D}$ should be chosen sufficiently large so that the truncation errors are negligible.

Numerical experiments show that the unboundedness of the quantum pressure term — particularly near the vortex core — can lead to numerical instability or even breakdown of the simulation.
A regularization of the quantum pressure term is therefore needed. 
In practice, we introduce a small regularization parameter $\varepsilon>0$ and solve the following regularized problem instead
\bea\label{DNLSE_regularized}
&&\p_t \phi(\bx, t) =  \left[\fl{1}{2}\nabla^2 -V(\bx) - \bt |\phi|^{\frac{4}{3}}
 - \ap\fl{\nabla^2|\phi|}{\sqrt{|\phi|^2+\varepsilon}} + \Og L_z + \mu \right]\phi(\bx, t), \quad \bx\in{\mathbb R}^d, \ \
t\in[t_n, t_{n+1}],\qquad\quad \, \\
\label{Projection_reg}
&&\phi(\bx, t_{n+1}) = \fl{\phi(\bx, t_{n+1}^-)}{\|\phi(\cdot, t_{n+1}^-)\|}, \quad \ \bx\in{\mathbb R}^d,
\quad\ \, \mbox{with}\quad\|\phi(\cdot, t_{n+1}^-)\|= \sqrt{\int_{{\mathbb R}^d}|\phi(\bx, t_{n+1}^-)|^2 d\bx},
\eea

From $t=t_n$ to $t=t_{n+1}$,  the system \eqref{DNLSE_regularized}–\eqref{Projection_reg} can be discretized using either a finite difference method or a pseudospectral method.
For simplicity, we present the finite difference discretization on a two-dimensional domain ${\mathcal D} = [a, b] \times [c, d]$. 
Its extension to other dimensions is straightforward.
Let $J,K>0$ be two integers, and define the mesh sizes $\Delta x=(b-a)/J$ and $\Delta y=(d-c)/K$. 
The grid points are $x_j=a+j\Delta x$ $(0\le j\le J)$ and $y_k=c+k\Delta y$ $(0\le k\le K)$.
Let $\tilde\phi^{\,n+1}_{j,k}$ and $\phi^n_{j,k}$ be the numerical approximations of $\phi(x_j,y_k,t_{n+1}^-)$ and $\phi(x_j,y_k,t_n)$ respectively, and set $V_{j,k} = V(x_j,y_k)$. 
We further denote by $\tilde{\Phi}^n = (\tilde{\phi}_{j,k}^n)$ and $\Phi^n = (\phi_{j,k}^n)$ the corresponding collections of grid values. 
Then a fully discrete semi-implicit regularized GFDN scheme is given by 
\begin{align}\label{scheme:GF}
&\frac{\tilde\phi^{\,n+1}_{j,k}-\phi^{n}_{j,k}}{\Delta t}
=\Big[\tfrac12\nabla_h^2 - V_{j,k}
-\beta\,|\phi^{n}_{j,k}|^{\frac{4}{3}}
-\alpha\,\tfrac{\nabla_h^2|\phi_{j,k}^{n}|}{\sqrt{|\phi^{n}_{j,k}|^{2}+\varepsilon}}
+\Omega\,L_{j,k} + \mu^n\Big]\tilde{\phi}^{n+1}_{j,k},\\
&\phi^{n+1}_{j,k}=\tilde\phi^{\,n+1}_{j,k}\Big/\|\tilde\Phi^{\,n+1}\|_h,\qquad 
\text{ with } \quad 
\|\tilde\Phi^{\,n+1}\|_h=\sqrt{\Delta x\,\Delta y\sum_{j,k}\big|\tilde\phi^{\,n+1}_{j,k}\big|^2}, \label{scheme:norm}
\end{align}
where $\nabla_h^2=\delta_x^2+\delta_y^2$, $L_{j,k}=-i\,(x_j\delta_y-y_k\delta_x)$,  
with $\delta_x^2\phi^n_{j,k}=(\phi^n_{j-1,k}-2\phi^n_{j,k}+\phi^n_{j+1,k})/(\Delta x)^2$,
$\delta_x\phi^n_{j,k}=(\phi^n_{j+1,k}-\phi^n_{j-1,k})/(2\Delta x)$,  and with analogous definitions in the $y$-direction. 
The discrete Lagrange multiplier $\mu^n$ can be computed as 
$$
\mu^n = \Delta x \Delta y\sum_{j,k} \left[-\fl{1}{2} \bar{\phi}_{j,k}^n \nabla_h^2\phi^n_{j,k}+ V_{j,k} |\phi_{j,k}^n|^2 +
\bt|\phi_{j,k}^n|^{10/3} + \ap \fl{|\phi_{j,k}^n|^2}{\sqrt{|\phi_{j,k}^n|^2+\varepsilon}} \nabla_h^2|\phi_{j,k}^n| - \Og{\rm Re}(\bar{\phi}_{j,k}^n L_{j,k}\phi_{j,k}^n)\right].
$$

\begin{remark}
When the external potential is radially symmetric, the regularized problem \eqref{DNLSE_regularized}–\eqref{Projection_reg} can be reformulated in polar coordinates in 2D, and in cylindrical coordinates in 3D.
A corresponding numerical scheme can then be derived in a manner analogous to that in \cite{Bao2013}.
\end{remark}
\begin{remark}
The above method can also be used to compute ground states of nonrotating/rotating Bose-Einstein condensates.
\end{remark}

\section{Numerical results}
\label{section5}

In this section, we numerically investigate the ground states of unitary Fermi gases with and without the angular-momentum rotation term.
Throughout this section, the ground states are computed using the regularized GFDN method \eqref{scheme:GF}–\eqref{scheme:norm}, and the iteration is terminated once $\|\Phi^{n+1}- \Phi^n \|_h < 10^{-6}$.

\subsection{For nonrotating case (i.e. $\Og = 0$)}
\label{section5-1}


{\bf Example 1 \  }  We study  the one-dimensional ground states of nonrotating unitary Fermi
gases, i.e., $d = 1$ and $\Og = 0$ in (\ref{DNLSE}).
We consider two types of external potentials: (i)  the harmonic potential $V_1(x) = \fl{1}{2}x^2$, and (ii)
the optical lattice potential $V_2(x) = \fl{1}{2}x^2+5\sqrt{2}\sin^2\left(\fl{\pi x}{2}\right)$.
The computational domain is chosen as ${\mathcal D} = [-20, 20]$ with mesh size
$\Delta x = 0.01$, and the time step is set to be $\Delta t =0.01$.

\begin{figure}[h!]
\centerline{
a)  \includegraphics[height=4.86cm,width=7.2cm]{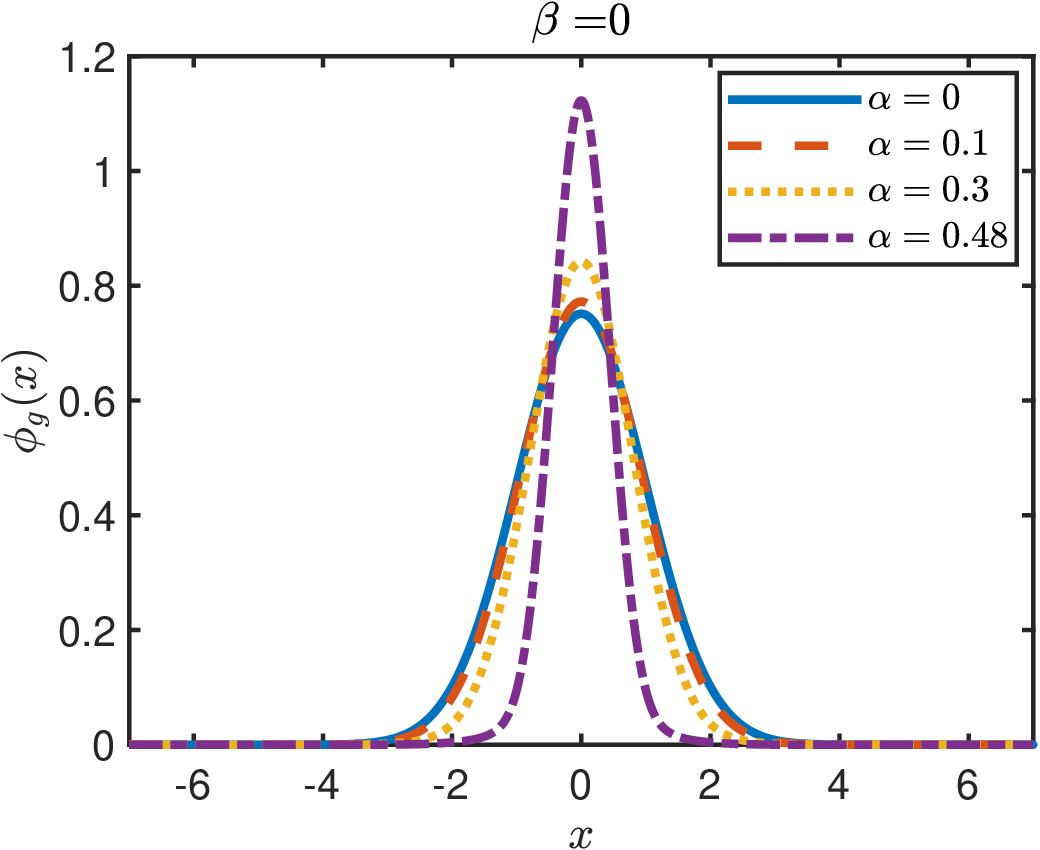}
b)  \includegraphics[height=4.86cm,width=7.2cm]{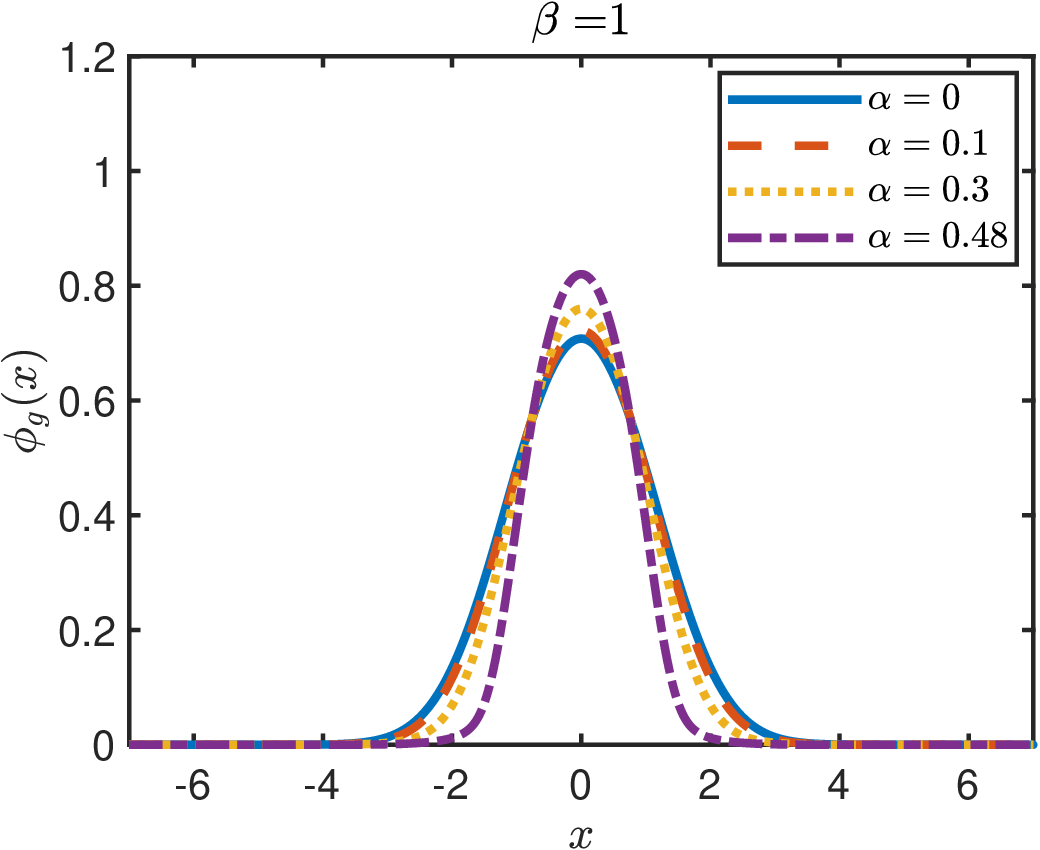}
 }
 \centerline{
c) \includegraphics[height=4.86cm,width=7.2cm]{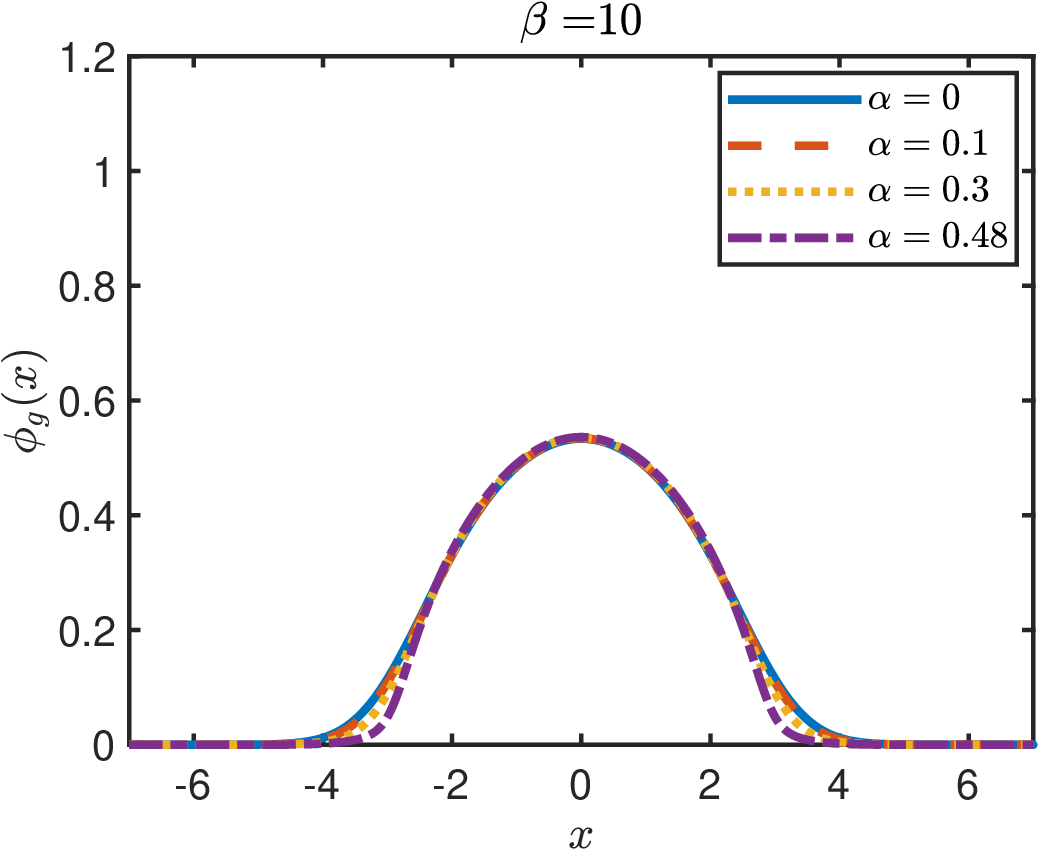}
d) \includegraphics[height=4.86cm,width=7.2cm]{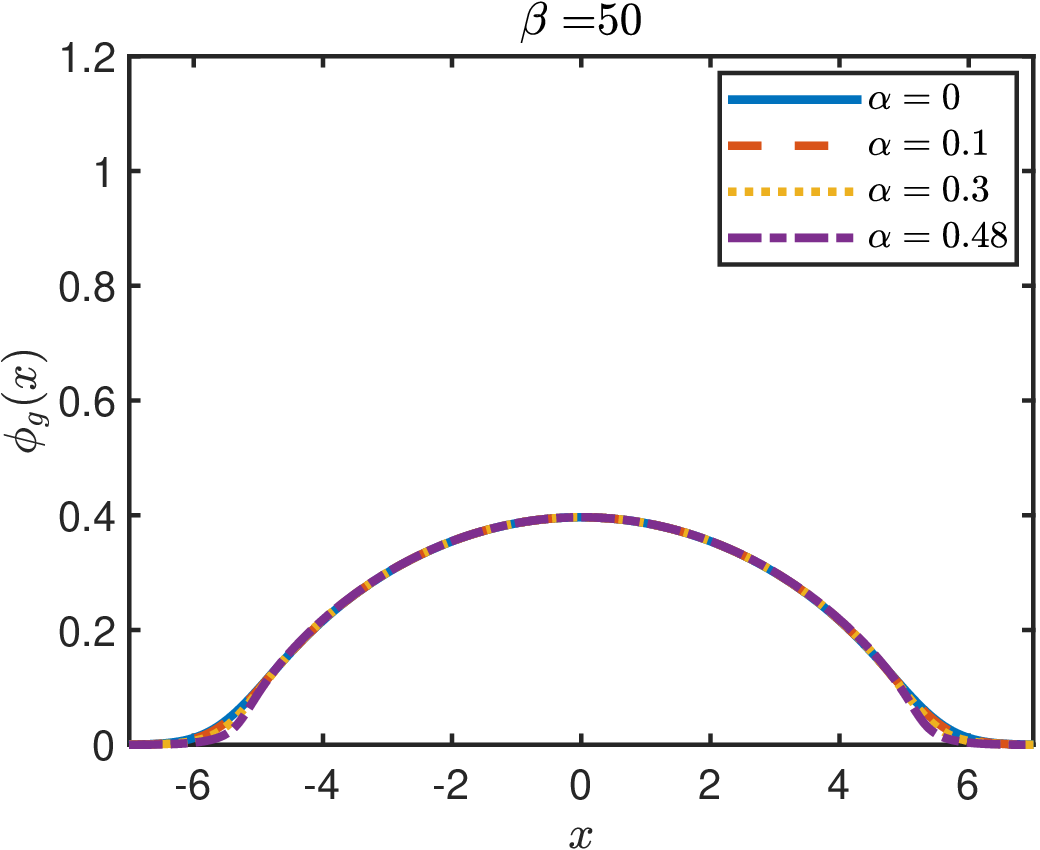}
 }
\caption{Ground states in the one dimensional harmonic potential $V_1(x) = \fl{1}{2}x^2$.}\label{F1}
\end{figure}

\begin{figure}[h!]
\centerline{
a) \includegraphics[height=4.86cm,width=7.2cm]{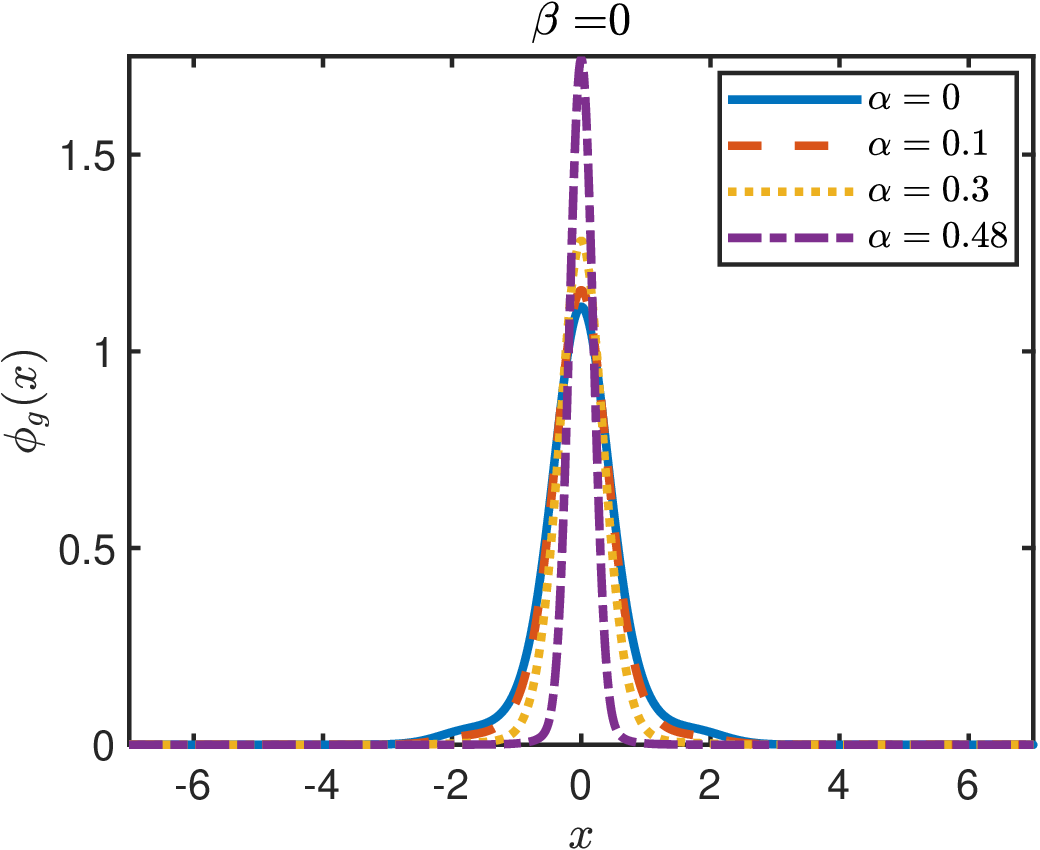}
b)  \includegraphics[height=4.86cm,width=7.2cm]{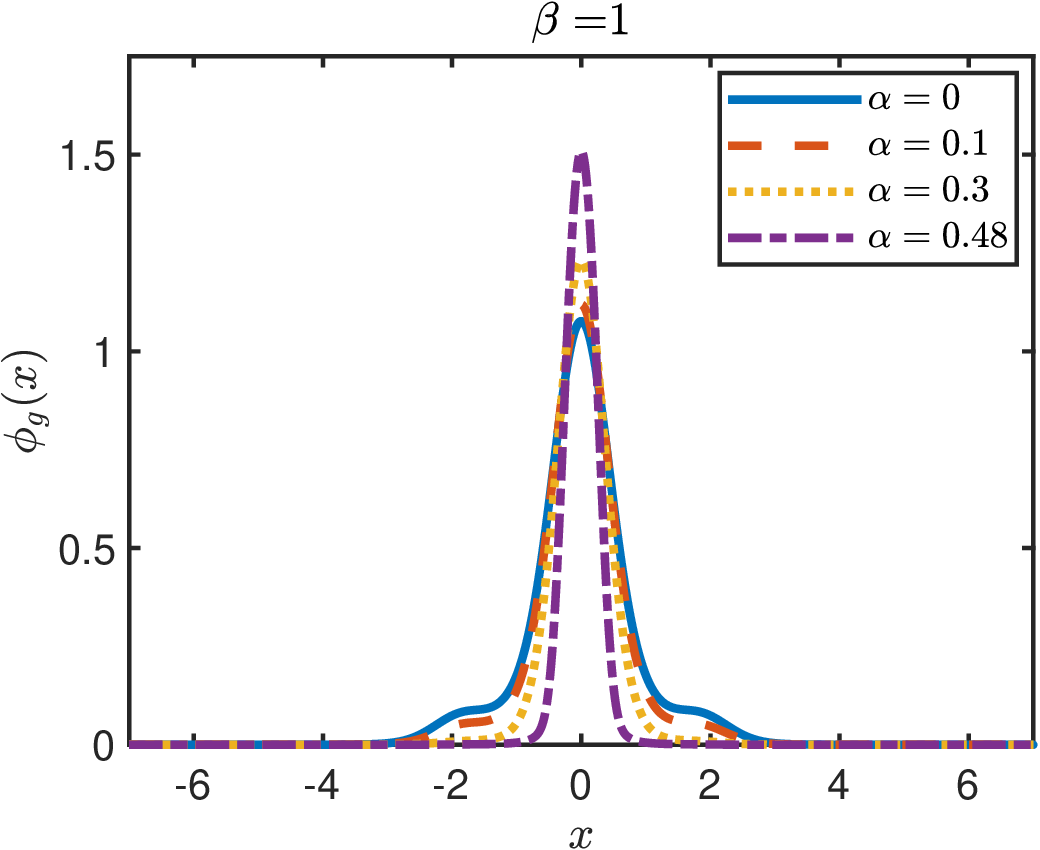}
 }
 \centerline{
c) \includegraphics[height=4.86cm,width=7.2cm]{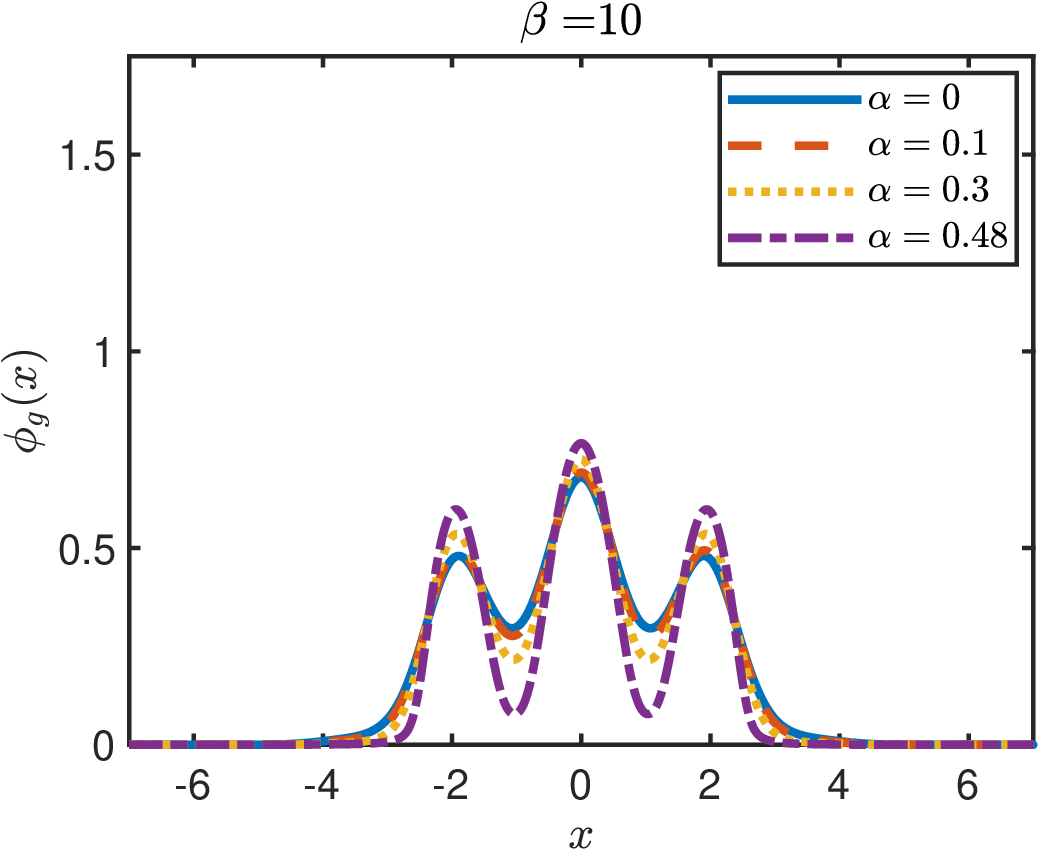}
d)  \includegraphics[height=4.86cm,width=7.2cm]{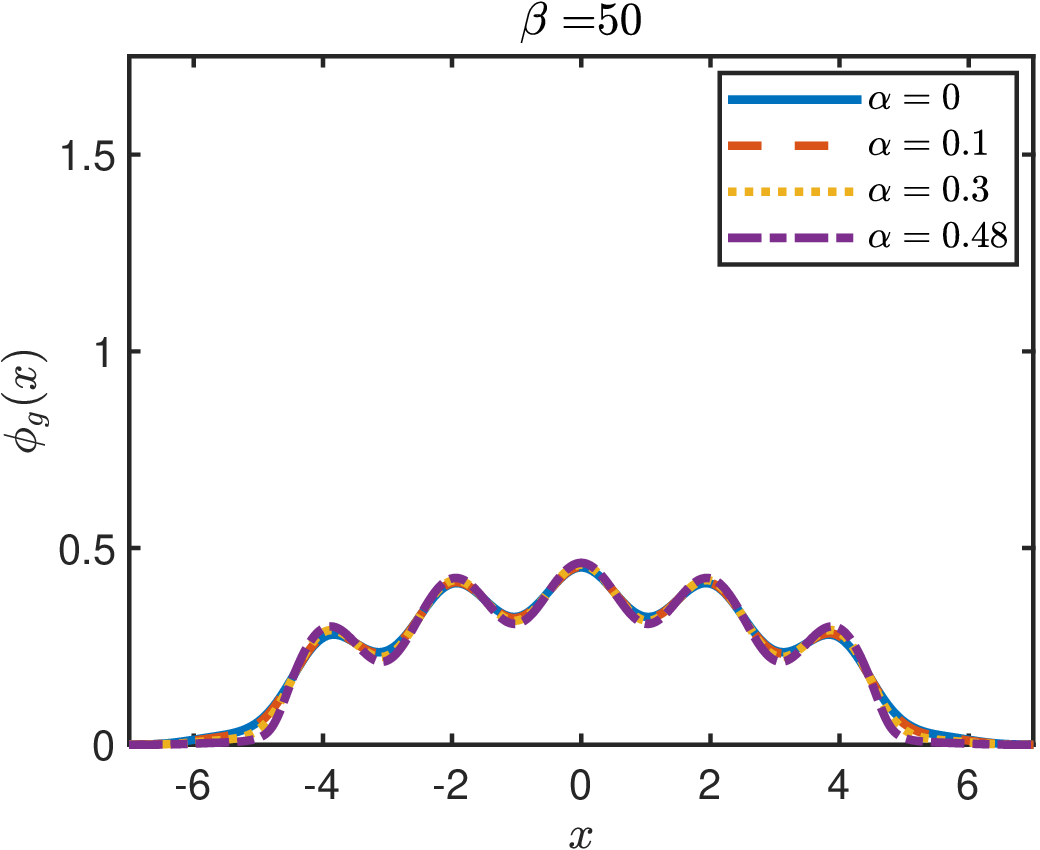}
 }
\caption{Ground states in the one dimensional optical lattice potential 
$V_2(x) = \fl{1}{2}x^2+5\sqrt{2}\sin^2\left(\fl{\pi x}{2}\right)$.}\label{F2}
\end{figure}

Fig. \ref{F1} and Fig. \ref{F2} show the positive ground states $\phi_g(x)$ under the harmonic
potential $V_1(x)$ and the optical lattice potential $V_2(x)$, respectively.  
We see that when $\bt$ increases, the peak value $\phi_g(0)$  decreases and the radius increases.
This behavior is similar to that observed in the ground states of nonrotating Bose–Einstein condensates \cite{Bao2004}.
On the other hand, we find that the effect of the quantum-pressure term is significant when $\bt$ is small.
Taking the harmonic potential as an example, Figures~\ref{F1}a–b show that, for small $\bt$,  increasing the parameter $\ap$,
leads to a narrower profile of $\phi_g(x)$.  
In contrast, when $\bt$ is large,
the influence of the quantum-pressure term becomes negligible, as shown in Fig. \ref{F1} d). 
In this regime, the Thomas–Fermi approximation \eqref{TFgs}, which neglects the kinetic and quantum pressure energies, provides an accurate description of the ground state. 
Similar behavior can be observed in the case of the optical lattices potential, as shown in Fig. \ref{F2}.

To further study the effect of quantum pressure term, we present the ground state energies for different $\bt$ and $\ap$ in Table \ref{T1}.  
Here $E_1(\phi_g)$ and $E_2(\phi_g)$ denote the ground state energies corresponding to the harmonic potential and the optical lattice potential, respectively.  
The results show that the relative difference between these energies is significant only when $\bt$ is small.
In other words, the effect of the quantum-pressure term is appreciable only in the small $\bt$ regime.
This observation is consistent with the behavior of the ground state profiles shown in Fig. \ref{F1} and Fig. \ref{F2}. 
These findings agree with those reported in \cite{Manini2005}.

\begin{table}[h!]
\begin{center}
\begin{tabular}{|l|cccc|cccc|}
\hline
& \multicolumn{4}{c|}{Energy $E_1(\phi_g)$} & \multicolumn{4}{c|}{
Energy $E_2(\phi_g)$}\\
\hline
$\ap$ & $\bt = 0$ & $\bt = 1$ & $\bt = 10$ & $\bt = 500$ & $\bt = 0$ & $\bt = 1$ &
$\bt = 10$ & $\bt = 500$\\
\hline
$0$      & 0.5000&  0.8062& 3.0028& 54.6196& 2.6484& 3.1592& 5.8762& 58.0842\\
$0.1$   & 0.4472& 0.7633& 2.9851& 54.6181& 2.4085& 2.9501& 5.7876& 58.0817\\
$0.2$   & 0.3873& 0.7163& 2.9667& 54.6166 & 2.1220& 2.6994& 5.6812& 58.0792\\
$0.3$   & 0.3162& 0.6634& 2.9476& 54.6151& 1.7654& 2.3901& 5.5489& 58.0766\\
$0.4$   & 0.2237& 0.6013& 2.9273 & 54.6136& 1.2767& 1.9813& 5.3751& 58.0739\\
$0.48$ &  0.1002& 0.5382& 2.9097& 54.6124& 0.5872& 1.4762& 5.1796& 58.0717\\
\hline
\end{tabular}
\caption{Ground state energies in the one dimensional harmonic potential $V_1(x)$ and the optical lattice potential $V_2(x)$ for various values of $\bt$ and $\ap$.}
\label{T1}
\end{center}
\end{table}

\bigskip

\noindent{\bf Example 2 \ }
Here we compare the numerically computed three-dimensional ground states with the corresponding low-dimensional effective models derived from the 3D NLSE (\ref{ONLSE}).  
We consider the external potential $V(\mathbf x) = (\gamma_x^2 x^2 + \gamma_y^2 y^2 + \gamma_z^2 z^2) /2$ and examine the following two configurations: 
(i) strong confinement in the transverse $(y,z)$-directions
($\gamma_y, \gamma_z \gg \gamma_x$),  leading to a  {\it cigar-shaped} condensate, 
and (ii) strong confinement along the axial direction 
($\gamma_z \gg \gamma_x, \gamma_y$), leading to a {\it disk-shaped} condensate.  
For both configurations, the corresponding effective lower dimensional models have been derived in Section \ref{section2}.

\begin{figure}[h!]
\centerline{
\includegraphics[height=3.86cm,width=5cm]{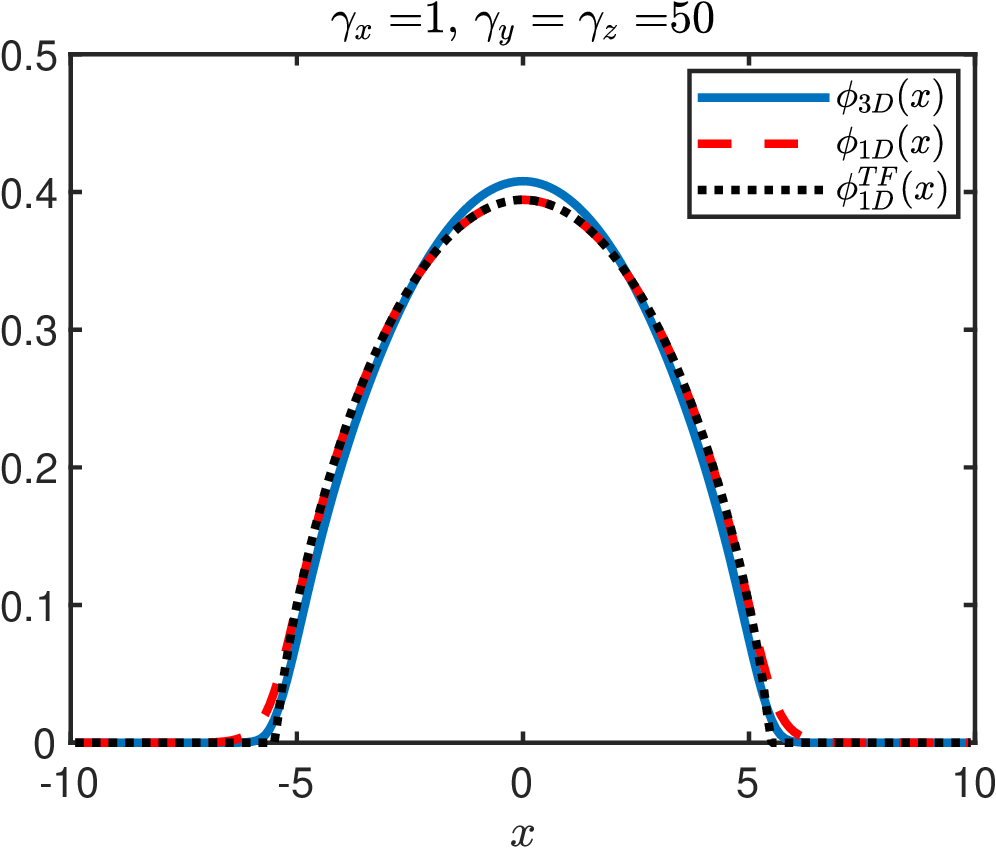}
\includegraphics[height=3.86cm,width=5cm]{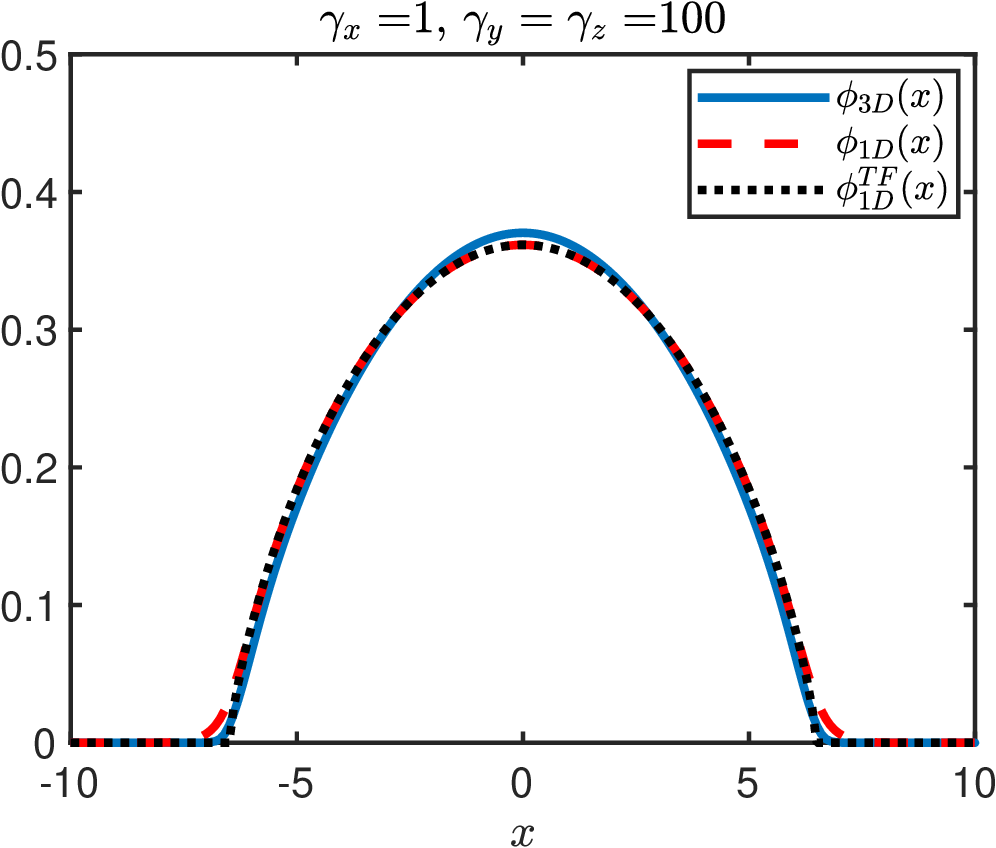}
\includegraphics[height=3.86cm,width=5cm]{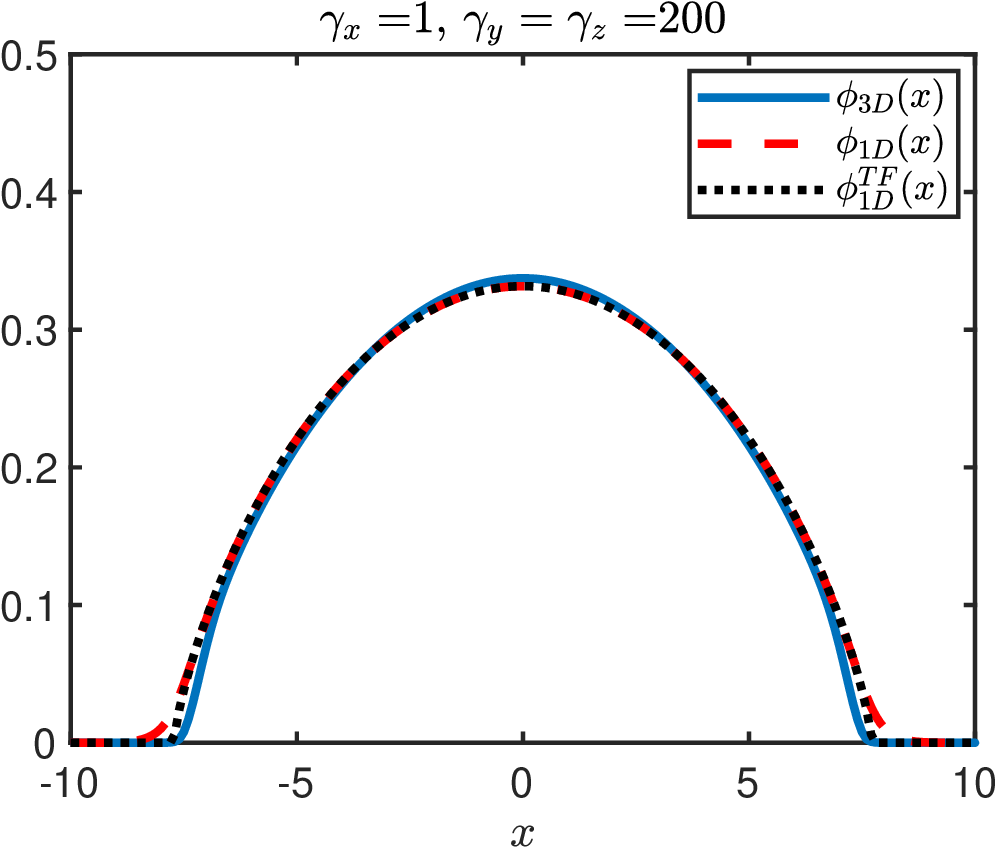}
 }
 \centerline{
\includegraphics[height=3.86cm,width=5cm]{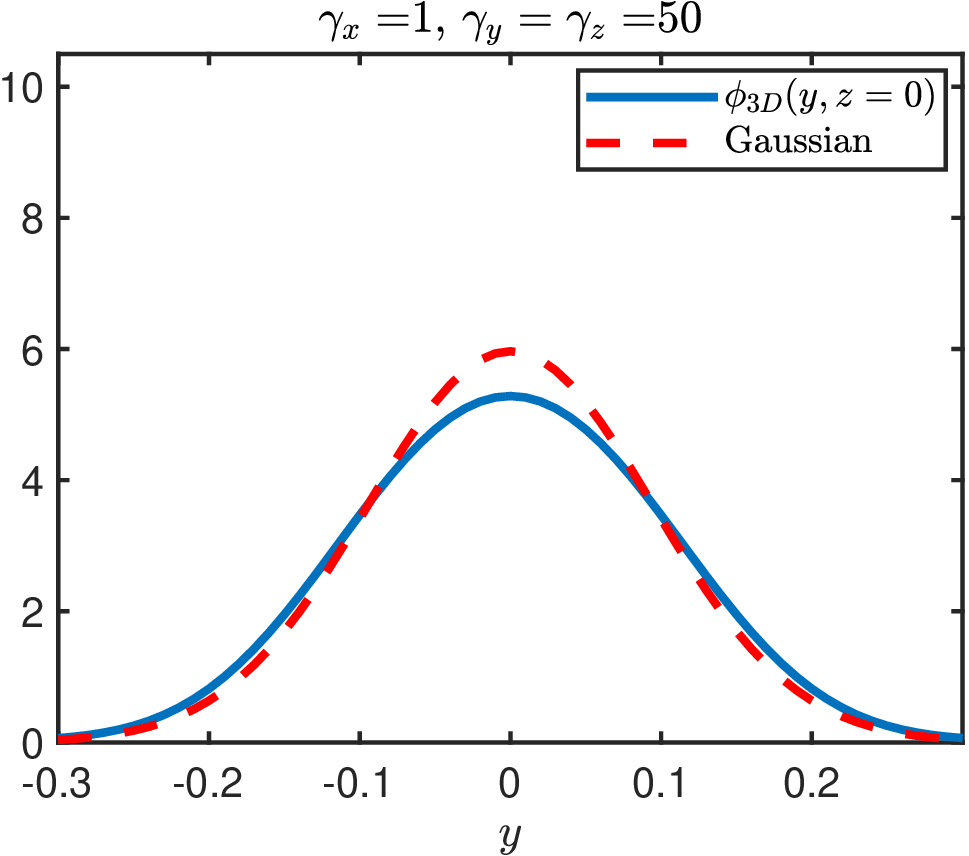}
\includegraphics[height=3.86cm,width=5cm]{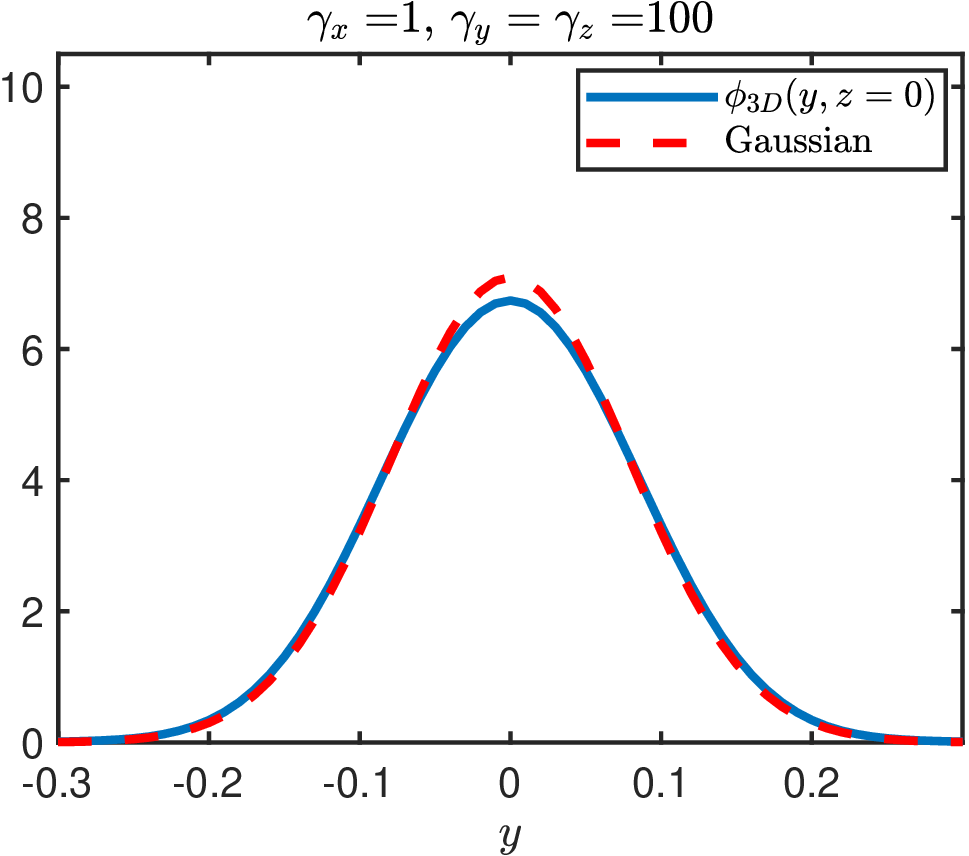}
\includegraphics[height=3.86cm,width=5cm]{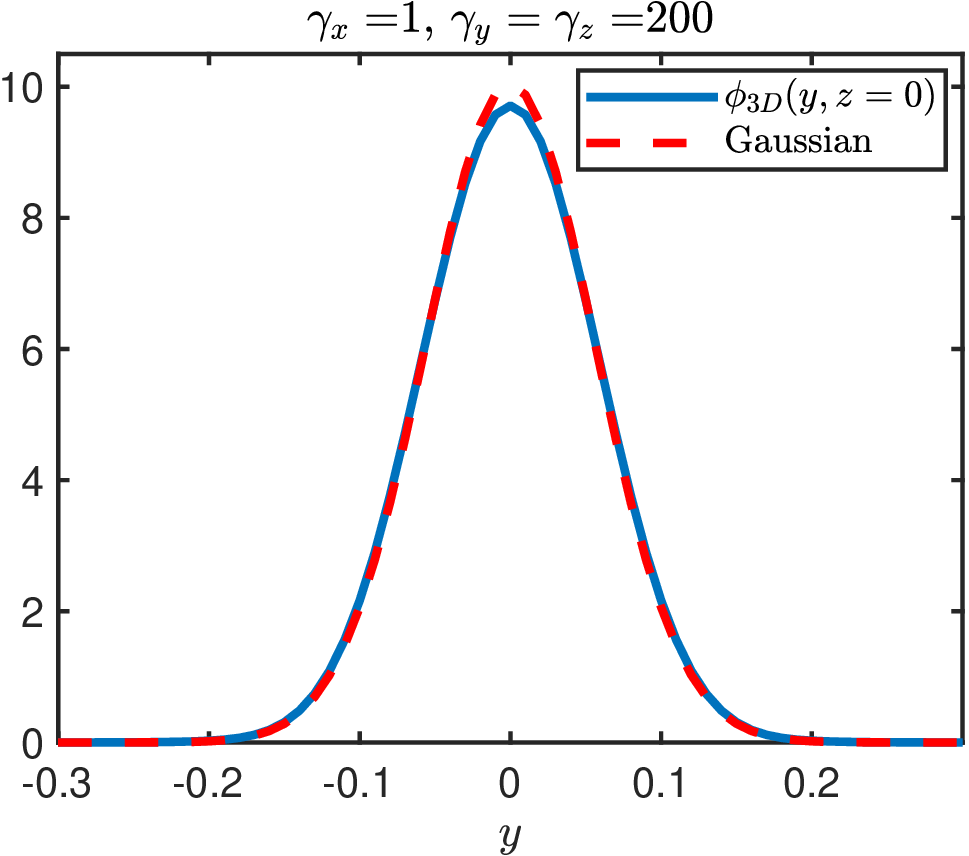}
 }
\caption{Numerical verification of the dimension reduction approximations for a cigar-shaped condensate under the harmonic potential $V(\bx) = (\gamma_x^2 x^2 + \gamma_y^2 y^2 + \gamma_z^2 z^2)/2$.  
Here we fix $\gamma_x = 1$ and set $\gamma_y = \gamma_z = 50$, 100, and 200, respectively.
The first row compares the normalized axial profile $\phi_{3D}(x) := \iint \phi(x,y,z) \, dydz / \| \iint \phi(x,y,z) \, dydz \|$, where $\phi(x,y,z)$ is the numerical ground state computed from the full 3D model, with the corresponding normalized solution $\phi_{1D}(x)$ obtained from the effective one-dimensional model~\eqref{gpe1d}, and with the Thomas–Fermi approximation $\phi_{1D}^{TF}(x)$ \eqref{TFgs}. 
The second row compares the normalized transverse profile $\phi_{\perp}(y,z=0) := \int \phi(x,y,z=0) \, dx / \| \int \phi(x,y,z) \, dx \|$
with the Gaussian ansatz \eqref{0gs}.}\label{Fig:DimReduction3to1}
\end{figure}

\begin{figure}[h!]
 \centerline{
\includegraphics[height=3.86cm,width=5cm]{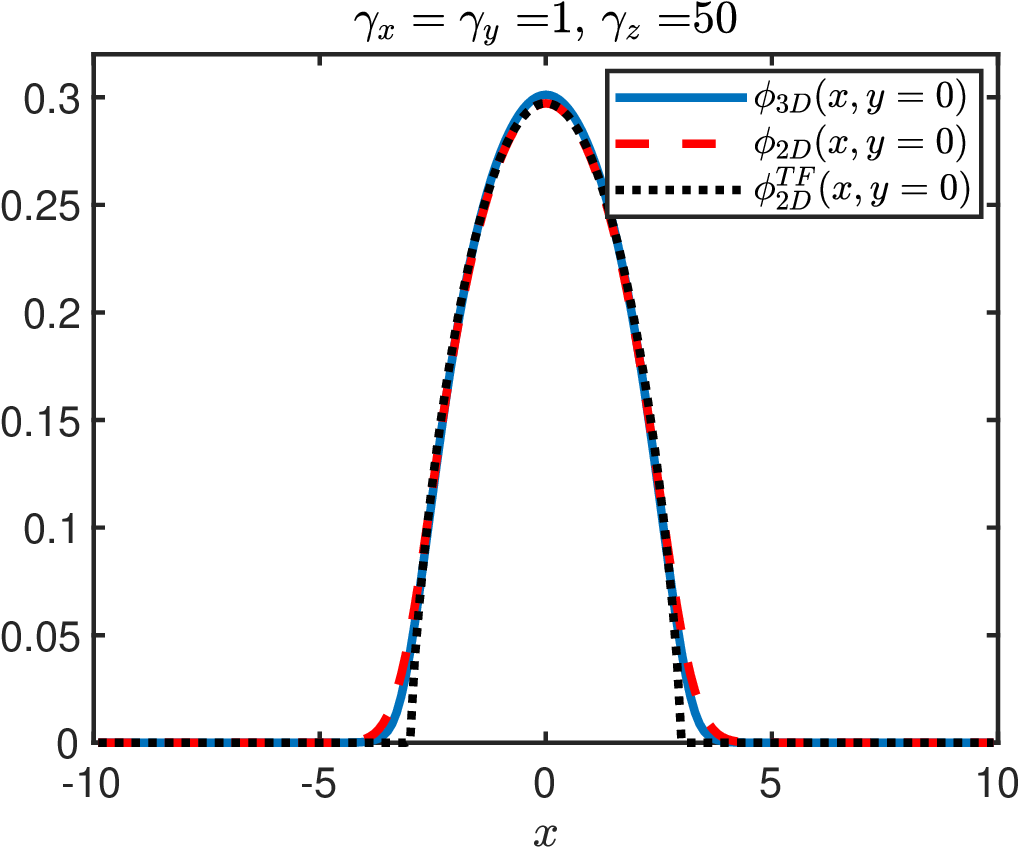}
\includegraphics[height=3.86cm,width=5cm]{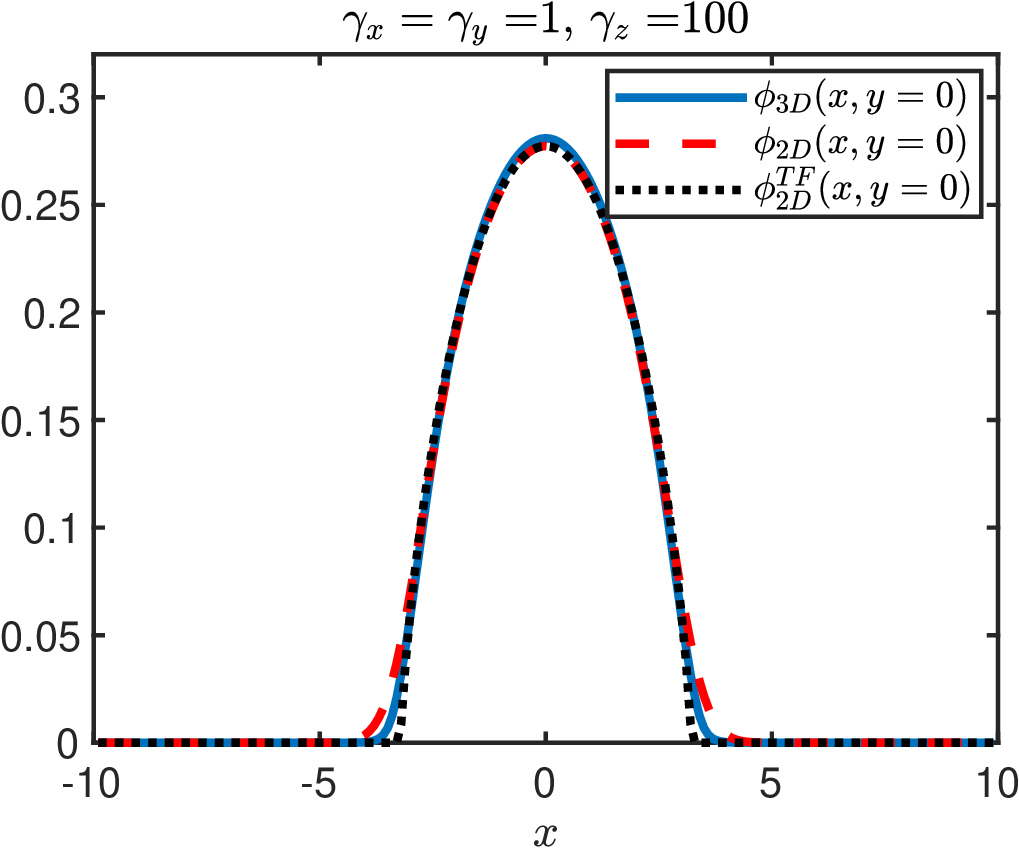}
\includegraphics[height=3.86cm,width=5cm]{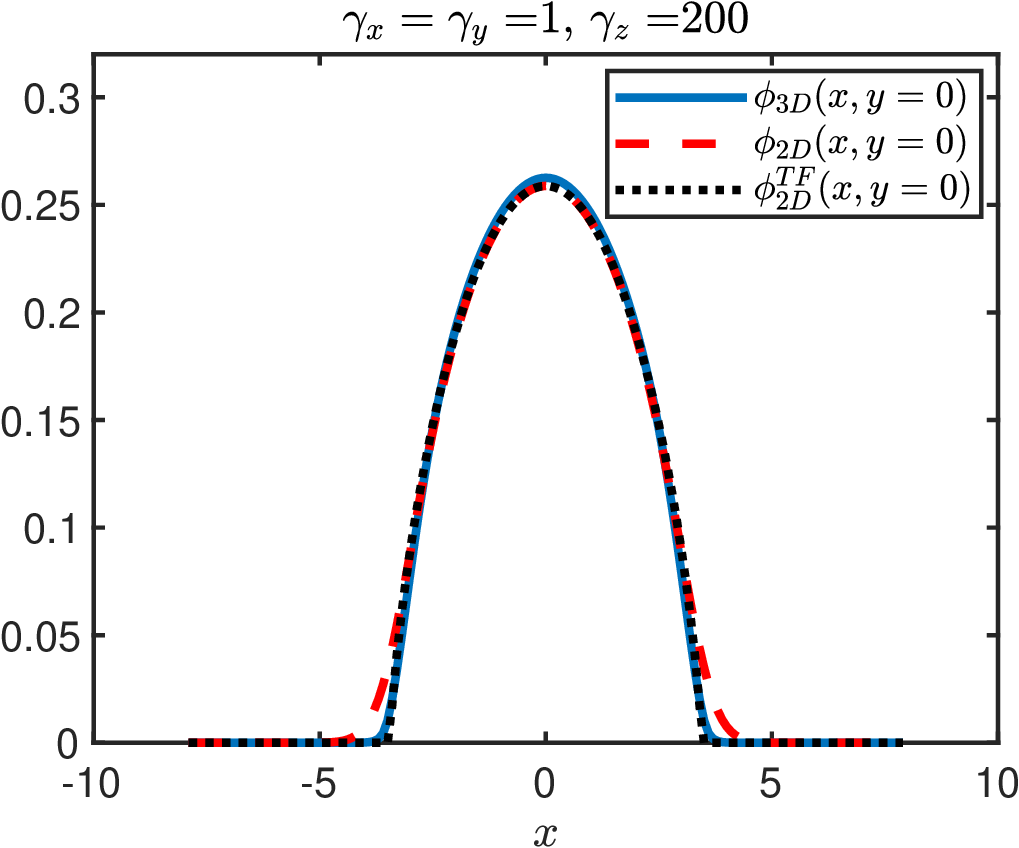}
 }
\centerline{
\includegraphics[height=3.86cm,width=5cm]{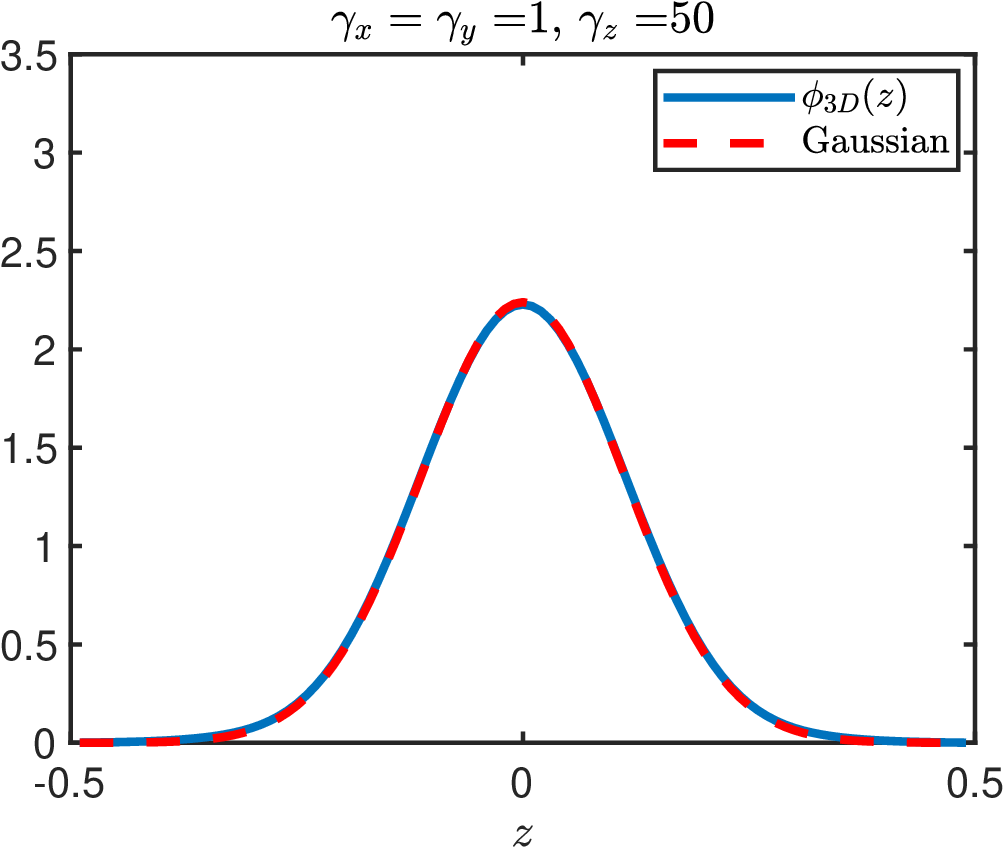}
\includegraphics[height=3.86cm,width=5cm]{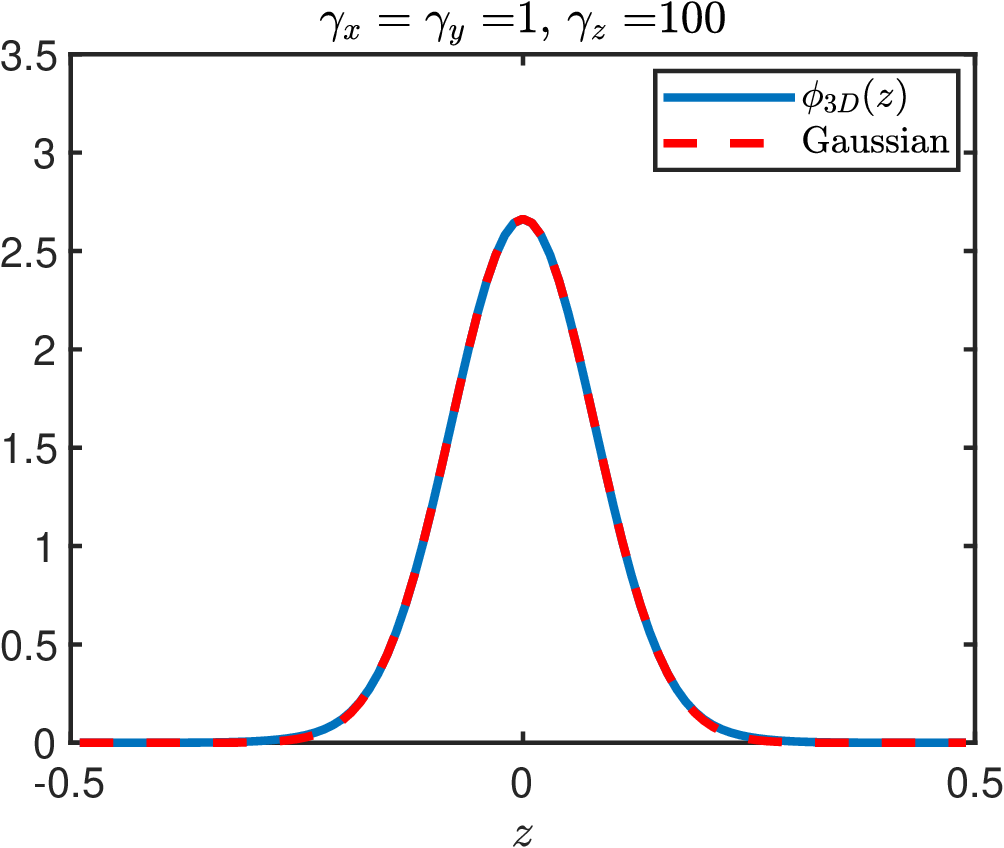}
\includegraphics[height=3.86cm,width=5cm]{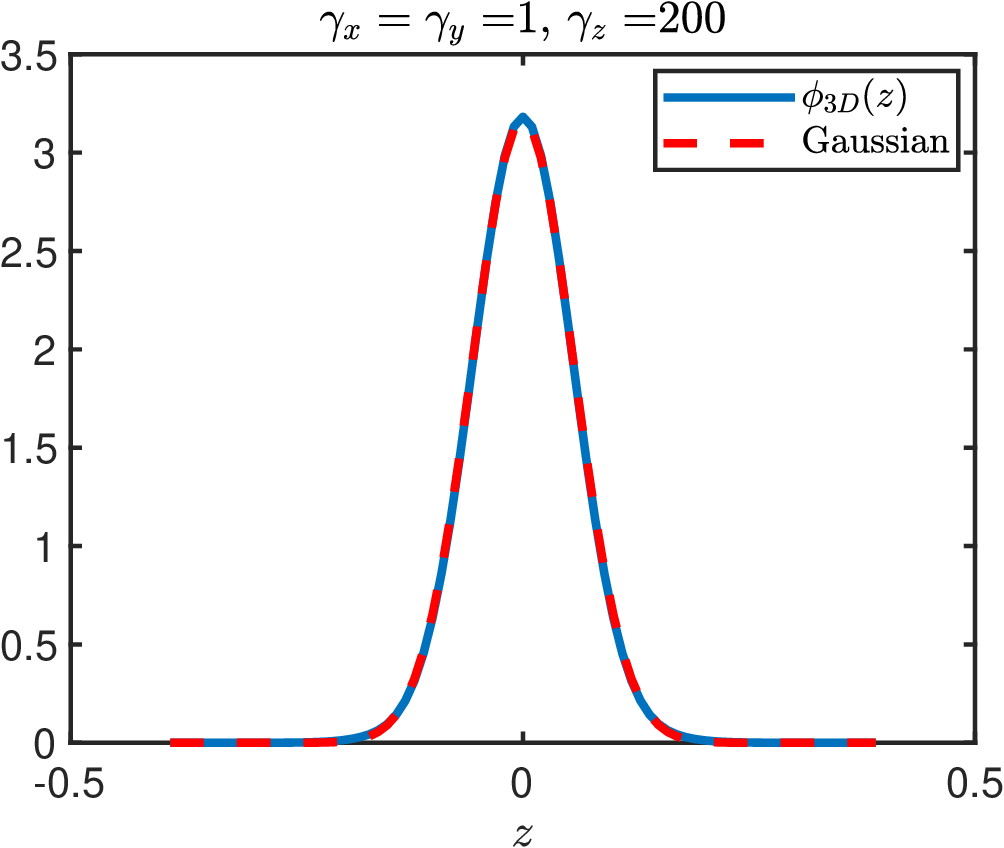}
 }
\caption{Numerical verification of the dimension reduction approximations for a disk-shaped condensate under the harmonic potential $V(\bx) = (\gamma_x^2 x^2 + \gamma_y^2 y^2 + \gamma_z^2 z^2)/2$.  Here we fix $\gamma_x = \gamma_y = 1$ and choose $\gamma_z = 50$, 100 and 200, respectively. 
The first row compares the normalized solution $\phi_{3D}(x, y=0) := \int \phi(x,y=0,z) \, dz / \| \int \phi(x,y,z) \, dz \|$, where $\phi(x,y,z)$ is the numerical solution computed from the 3D model, with the normalized numerical solution $\phi_{2D}(x, y=0)$, computed from the effective 2D model \eqref{gpe2d}, and with the Thomas-Fermi approximation $\phi_{2D}^{TF}(x, y=0)$ \eqref{TFgs}. The second row compares the normalized axial profile $\phi_{\perp}(z) := \int\phi(x,y,z) \, dxdy / \| \iint \phi(x,y,z) \, dx dy \|$
with the Gaussian ansatz \eqref{0gs}.}\label{Fig:DimReduction3to2}
\end{figure}

Fig.~\ref{Fig:DimReduction3to1} shows the comparison between the density profiles obtained from the full 3D model and the corresponding 1D effective model for a cigar-shaped case.  
The two results exhibit good agreement when the transverse trapping frequencies are sufficiently large, confirming the validity of the effective 1D model~\eqref{gpe1d} in the weakly confined direction and the Gaussian ansatz~\eqref{0gs} in the tightly confined directions. 
As shown in the figure, stronger confinement in the transverse $(y,z)$-directions leads to higher accuracy of the lower-dimensional model.
The results also demonstrate that the Thomas–Fermi approximation \eqref{TFgs} remains accurate when the interaction strength is large.
Similarly, Fig.~\ref{Fig:DimReduction3to2} validates the lower-dimensional model for the disk-shaped condensate, again showing good agreement between the reduced and the full 3D model.

\noindent{\bf Example 3 \ } We study the two-dimensional ground states and the central vortex
 states under the harmonic potential $V(x,y) = (x^2 +  y^2)/2$.
In 2D, when the external potential is radially symmetric, a central vortex state can be written in the form
\bea\label{vs}
\phi_{v}(r,\theta) =f_m(r) e^{im\tht}, \qquad \bx\in{\mathbb R}^2,
\eea
where $(r, \tht)$ is the polar coordinate,  $f_m(r)$ is a real-valued positive function and
$m\in{\mathbb Z}$ is usually called the index (or winding number) of the vortex state.  
A central vortex ground state is defined as the minimizer of the energy functional subject to the normalization constraint  $\|\phi_v(\bx)\| = 1$, together with the additional condition $\phi_v(0,\theta) = 0$, see \cite{Bao2013}. 

\begin{figure}[h!]
\centerline{
 \includegraphics[height=4.86cm,width=7.2cm]{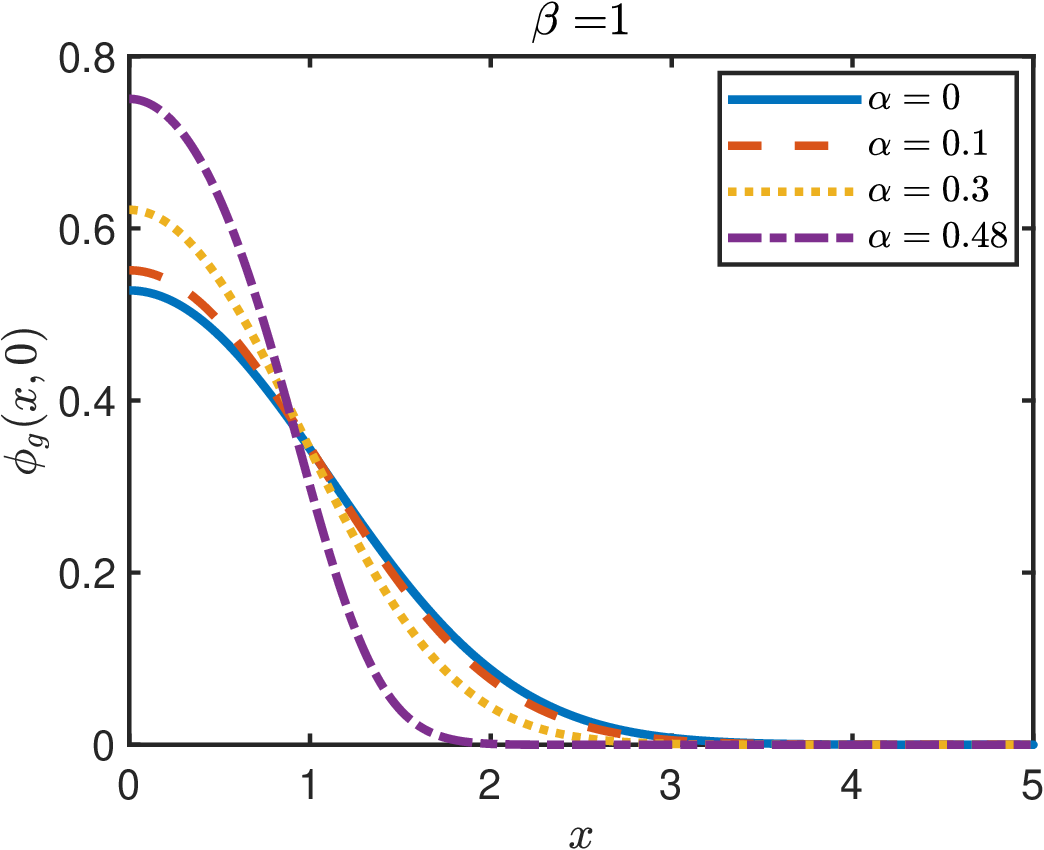}
 \includegraphics[height=4.86cm,width=7.2cm]{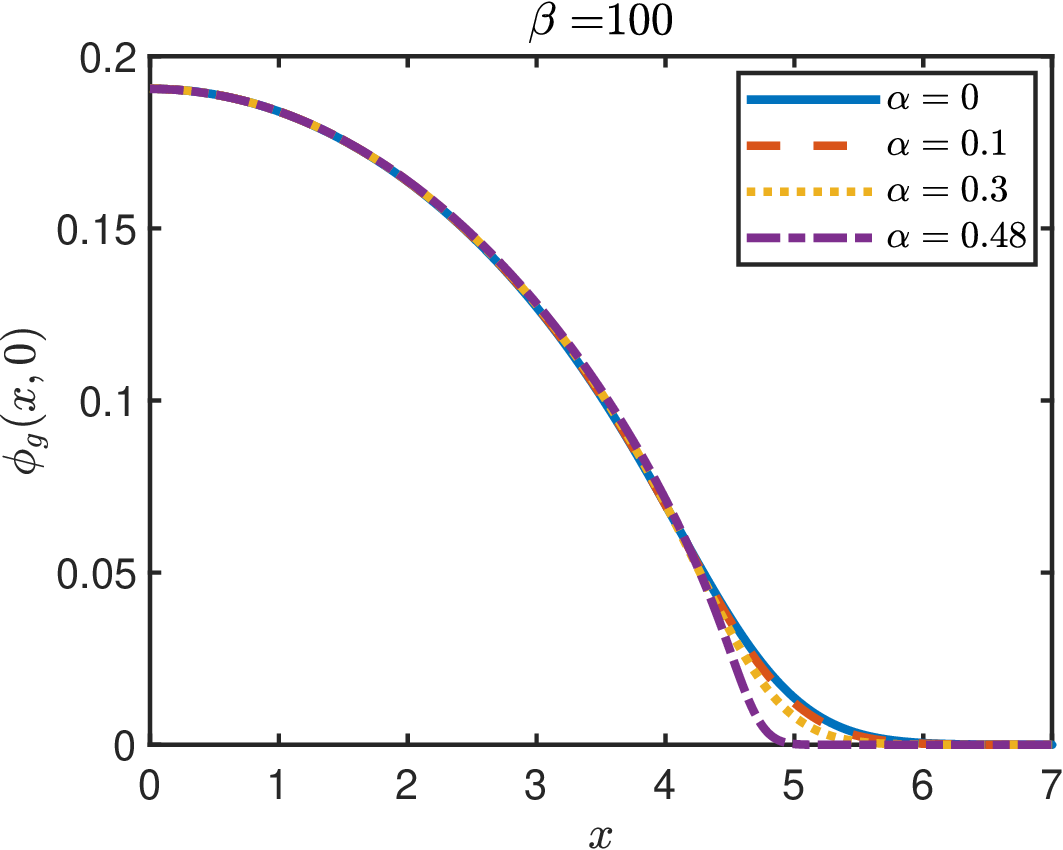}
 }
 \centerline{
 \includegraphics[height=4.86cm,width=7.2cm]{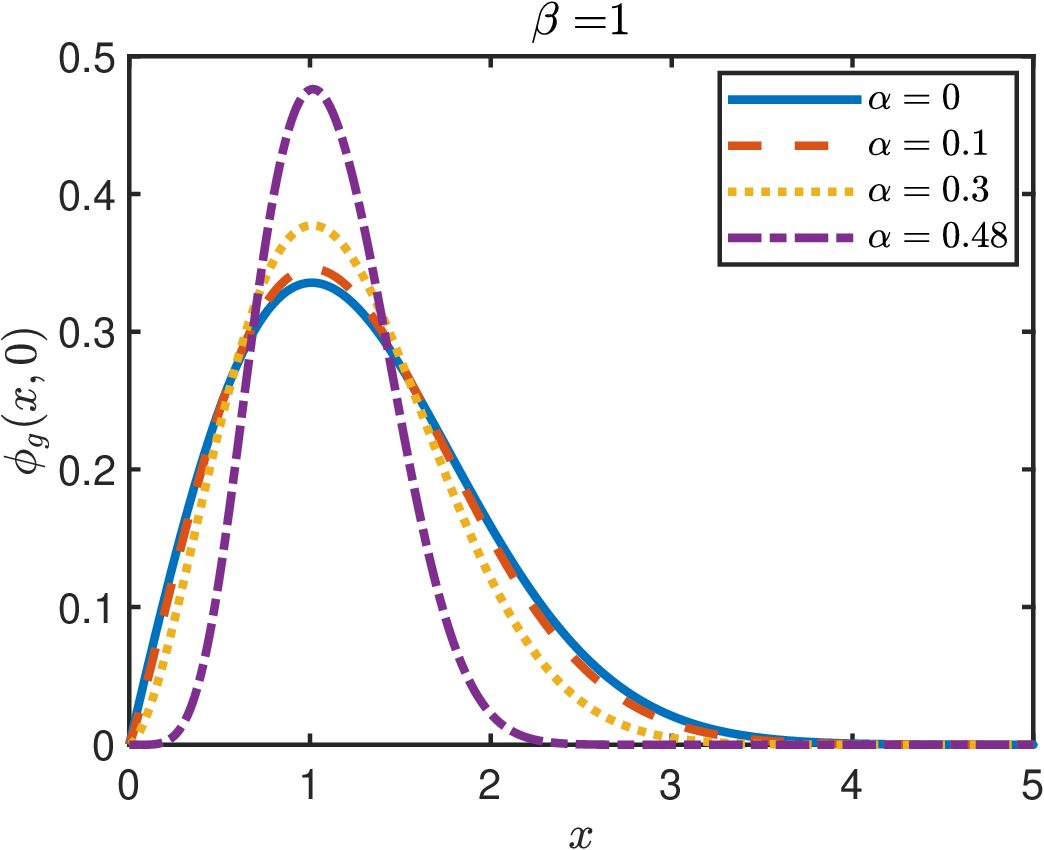}  
 \includegraphics[height=4.86cm,width=7.2cm]{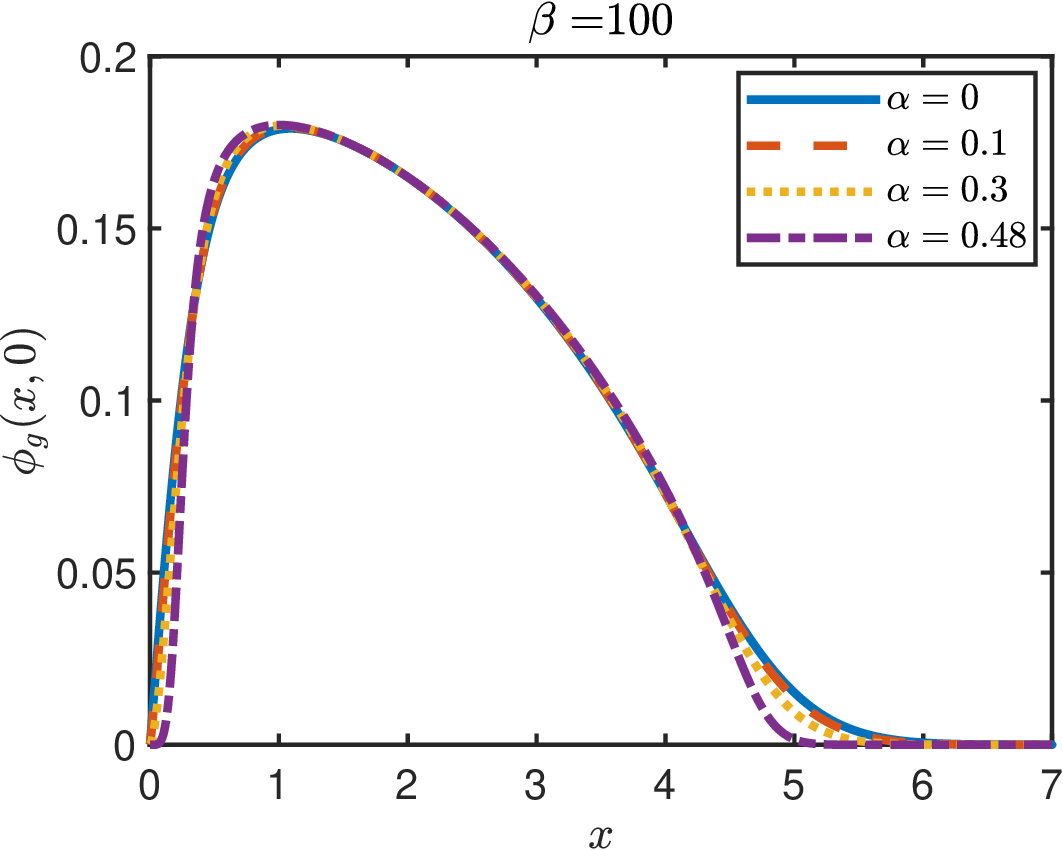}
 }
\caption{Ground states (top) and central vortex ground states with index $m = 1$
(bottom) in harmonic potential
$V(x,y) = \fl{1}{2}(x^2+y^2)$.}\label{F3}
\end{figure}

Fig. \ref{F3} shows the density profiles of the ground state and the central vortex ground state with $m = 1$.  
For ease of illustration, we plot only the values along the line $y = 0$ with $x > 0$, i.e. $\phi_g(x, 0)$ and $\phi_v(r,0) = f_1(r)$ in (\ref{vs}).
Table \ref{T2} lists the corresponding energies for different values of $\bt$ and $\ap$.  
The results show that the influence of the quantum pressure term is significant only when $\bt$ is small, which is consistent with the one dimensional observations in Example~1.
Moreover, we observe that the vortex core size increases as $\ap$ increases. 
Table \ref{T2} shows that, for the same parameters, the energy of the central vortex state is always higher than the corresponding ground state energy, which indicates that the central vortex state is an excited state. 
In addition, both $E_{\rm ho}(\phi_g)$ and $E_{\rm ho}(\phi_v)$ varies considerably with $\ap$ when $\bt$ is small, but changes only slightly when $\bt$  is large.
This again confirms that the quantum pressure term plays a significant role only in the small $\bt$ regime.

\begin{figure}[h!]
\centerline{
 \includegraphics[height=5.96cm,width=7.96cm]{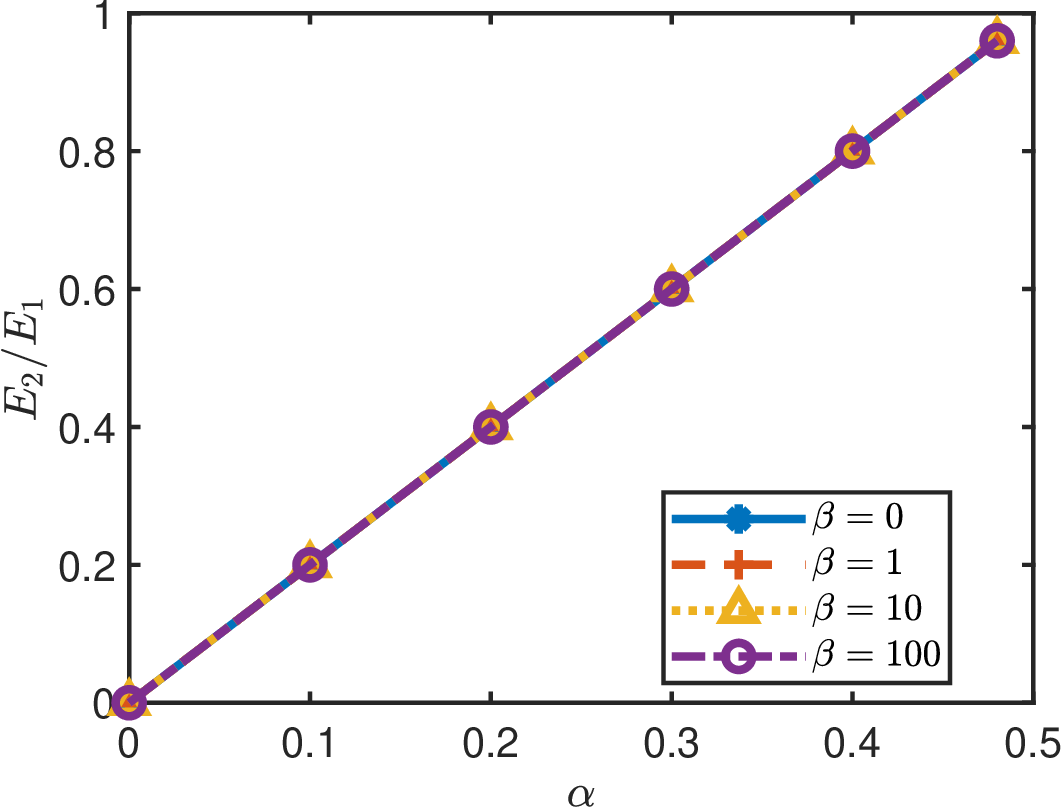}
 \includegraphics[height=5.96cm,width=7.96cm]{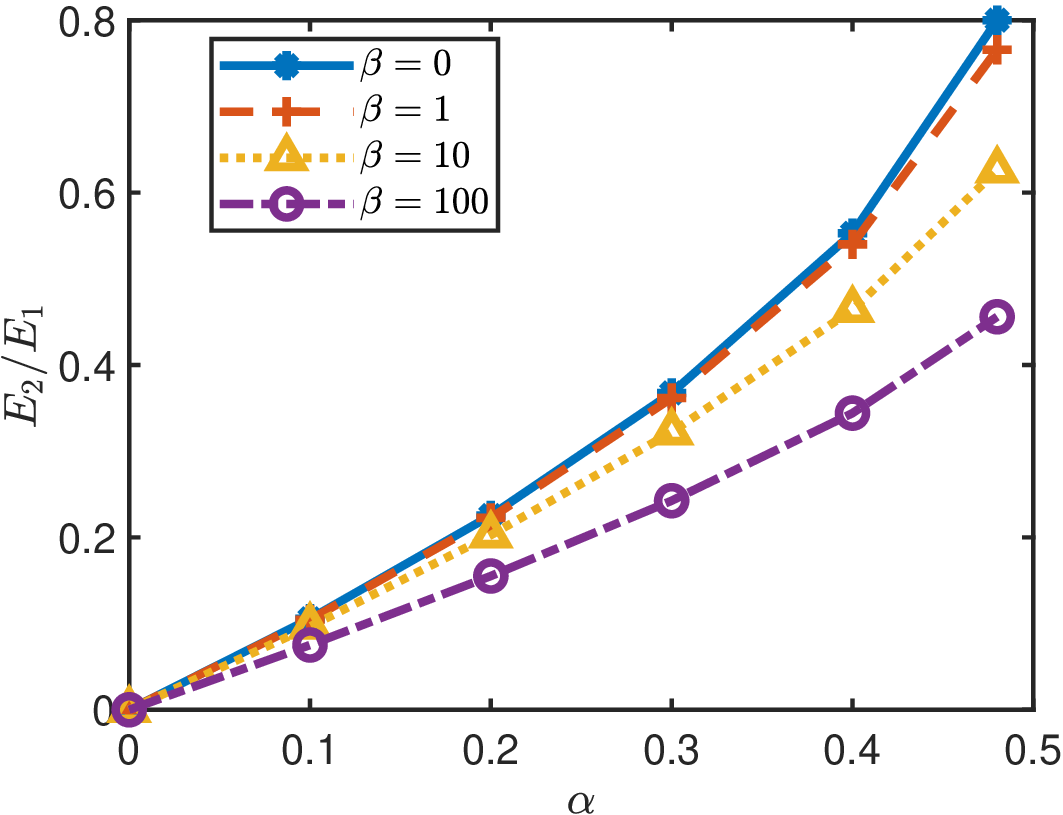}
}
\caption{Ratio $E_{2}/E_{1}$ versus parameter $\ap$ for the ground states
$\phi_g(\bx)$ (left) and central vortex states $\phi_{v}$ with $m=1$ (right).} \label{F4}
\end{figure}

To further examine the effects of the quantum-pressure term in ground states and central vortex states, we investigate the ratio $E_{2}/E_{1}$, where the quantities $E_1$ and $E_2$ are  defined as 
\bea
E_{\rm 1}(\phi) = \fl{1}{2}\int_{{\mathbb R}^2} |\nabla\phi(\bx)|^2 d\bx, \qquad
E_{\rm 2}(\phi) = \ap \int_{{\mathbb R}^2} |\nabla|\phi(\bx)||^2 d\bx,
\eea
which correspond to the first and the fourth term in the energy functional (\ref{energy}), respectively. 
As shown in Fig. \ref{F4}, when $\bt$ is fixed,  the ratio $E_2/E_1$ increases monotonically with respect to $\ap$.  
For ground states, this ratio is proportional to $\ap$,
namely $E_2/E_1 = 2\ap$. 
This behavior is expected since $|\nabla\phi_g(\bx)| = |\nabla|\phi_g(\bx)||$ for any $\bx\in{\mathbb R}^2$ and $\bt \geq 0$.  
In fact, when computing ground states of nonrotating unitary Fermi gases, one may solve the real-valued solution $\phi_{g}(\bx)$ from
\bea\label{eq:nonrotatingNLSE}
\p_t\phi(\bx,t) =  \left[\left(\frac12 - \ap\right)\nabla^2 - V(\bx) - \bt|\phi|^{4/3} + \mu\right]\phi(\bx,t), \quad
\ \bx\in{\mathbb R}^d,
\eea
instead of the equation (\ref{DNLSE}), 
and all ground states can be recovered in the form $ \phi_g(\bx) e^{i\theta}$ with arbitrary $\theta\in{\mathbb R}$.
In this setting, the quantum-pressure term can be absorbed into the Laplacian term, reducing the model to those studied in  \cite{Adhikari2009, Adhikari2009-1}.  
For central vortex states, the ratio increases as $\bt$ increases due to the nontrivial angular phase associated with vortex solutions. 
Besides, the ratio is always less than $2\ap$, confirming that $|\nabla|\phi_{v}||^2 < |\nabla\phi_{v}|^2$.  
The result implies that, when quantum vortices are present, the phase is non-uniform and the quantum pressure term must be retained explicitly in the NLSE \eqref{DNLSE}, which is in contrast to the equation \eqref{eq:nonrotatingNLSE}.

\begin{table}[htp!]
\begin{center}
\begin{tabular}{|l|cccc|cccc|}
\hline
& \multicolumn{4}{c|}{Energy $E_{\rm ho}(\phi_g)$} & \multicolumn{4}{c|}{
Energy $E_{\rm ho}(\phi_v)$}\\
\hline
$\ap$ & $\bt = 0$ & $\bt = 1$ & $\bt= 10$ & $\bt = 100$ & $\bt = 0$ & $\bt = 1$ &
$\bt = 10$ & $\bt = 100$\\
\hline
$0$      & 1.0000&1.1615&  2.2846&  7.9390&2.000&2.1064&2.9637&8.2490\\
$0.1$   & 0.8944&1.0671&  2.2295&  7.9189& 1.8944&2.0056&2.8912&8.2224\\
$0.2$   & 0.7746&0.9625&  2.1718&  7.8981& 1.7746&1.8921&2.8132&8.1948\\
$0.3$   & 0.6325&0.8432&  2.1107&  7.8768& 1.6324&1.7592&2.7274&8.1659\\
$0.4$   & 0.4472&0.6997&  2.0446&  7.8546& 1.4472&1.5905& 2.6300&8.1349\\
$0.48$ & 0.2000&0.5481 & 1.9848& 7.8357 & 1.2000& 1.3843& 2.5355&8.1071\\
\hline
\end{tabular}
\caption{Ground state energies $E(\phi_g)$ and central vortex energies $E(\phi_v)$ with $m = 1$, where a two-dimensional harmonic potential $V(x,y) = \fl{1}{2}(x^2 + y^2)$
is used.}
\label{T2}
\end{center}
\end{table}

 \subsection{For rotating case (i.e., $\Og \neq 0$)}
 \label{section5-2}

In the following, we study the two-dimensional ground states of rotating unitary Fermi gases.
We focus on the harmonic potential $V(x,y)=(x^2 + y^2)/2$. 
As shown in Theorem~\ref{thm:gs2}, ground states exist only when $\Omega < 1$. 
As $\Omega$ increases, quantum vortices start to appear in the ground state.
The existence of quantum vortices and the quantum pressure term makes the numerical computation of ground states highly challenging.
As discussed in Section~\ref{section4}, a regularization of the quantum-pressure term is required.
Throughout this section, we choose the regularization parameter $\varepsilon=10^{-4}$ in the scheme \eqref{scheme:GF}--\eqref{scheme:norm}.

\begin{figure}[h!]
 \centerline{
  \includegraphics[height=3.8cm,width=4.2cm]{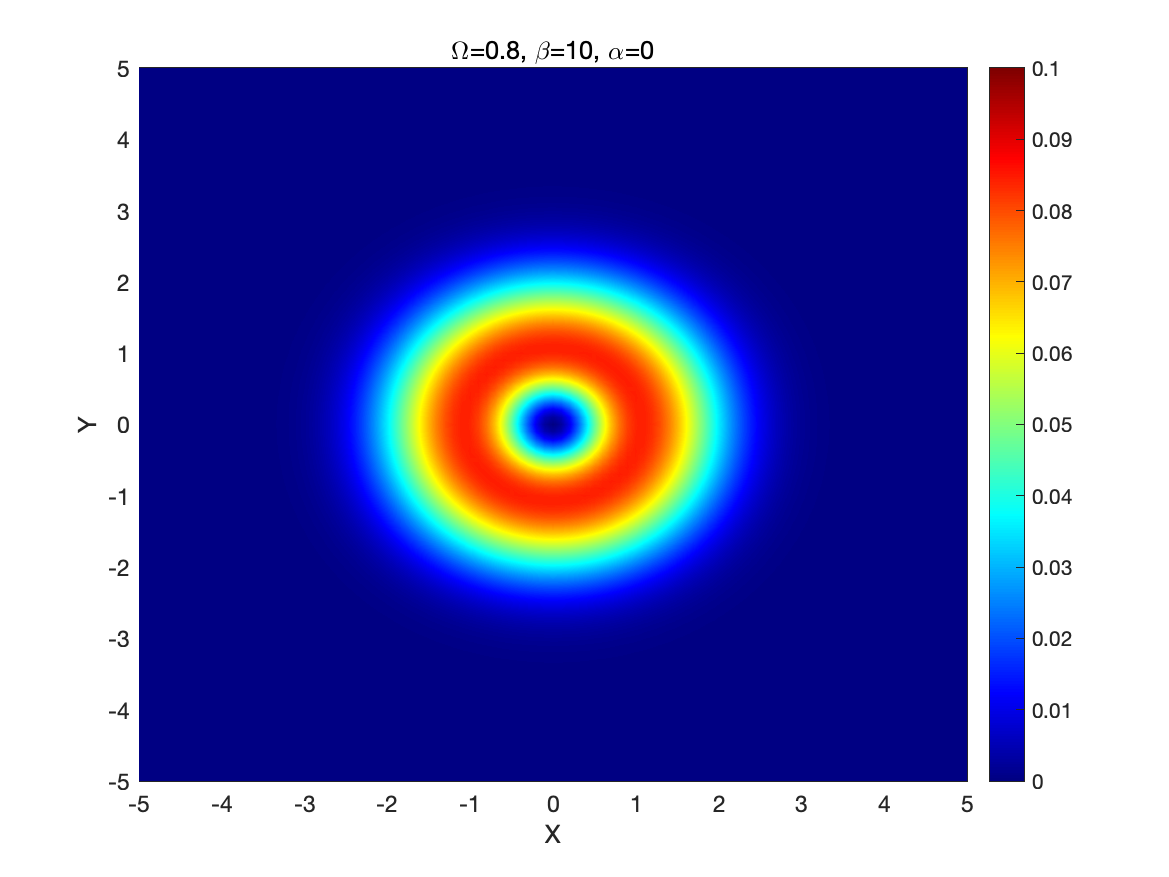}
\includegraphics[height=3.8cm,width=4.2cm]{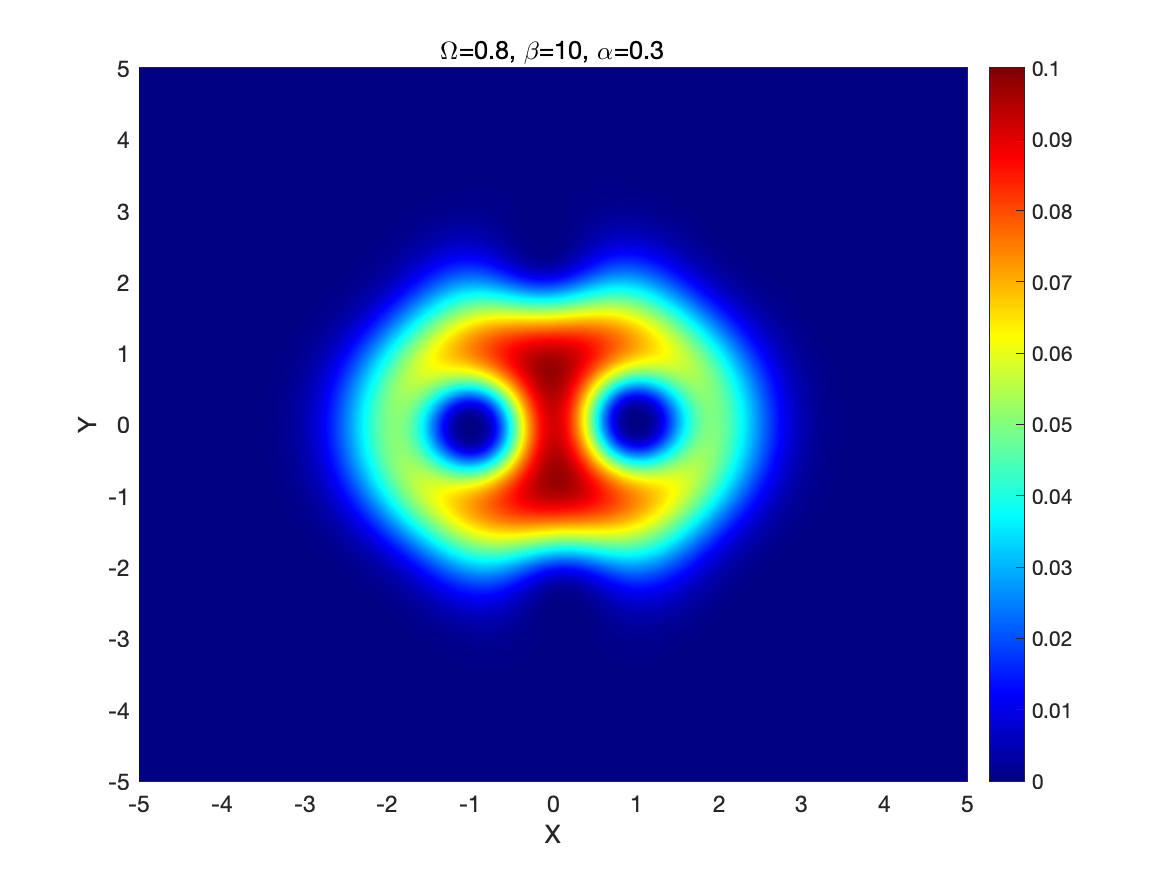}
 \includegraphics[height=3.8cm,width=4.2cm]{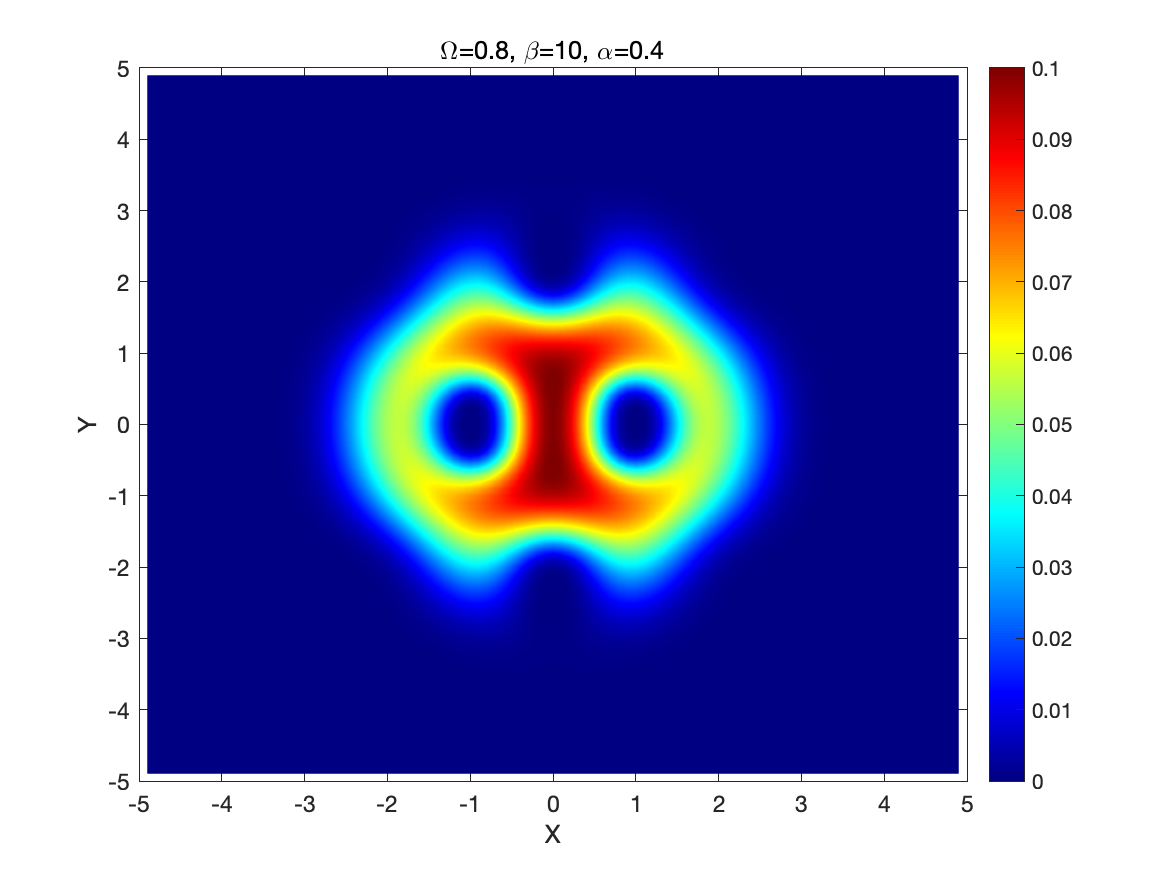}
 \includegraphics[height=3.8cm,width=4.2cm]{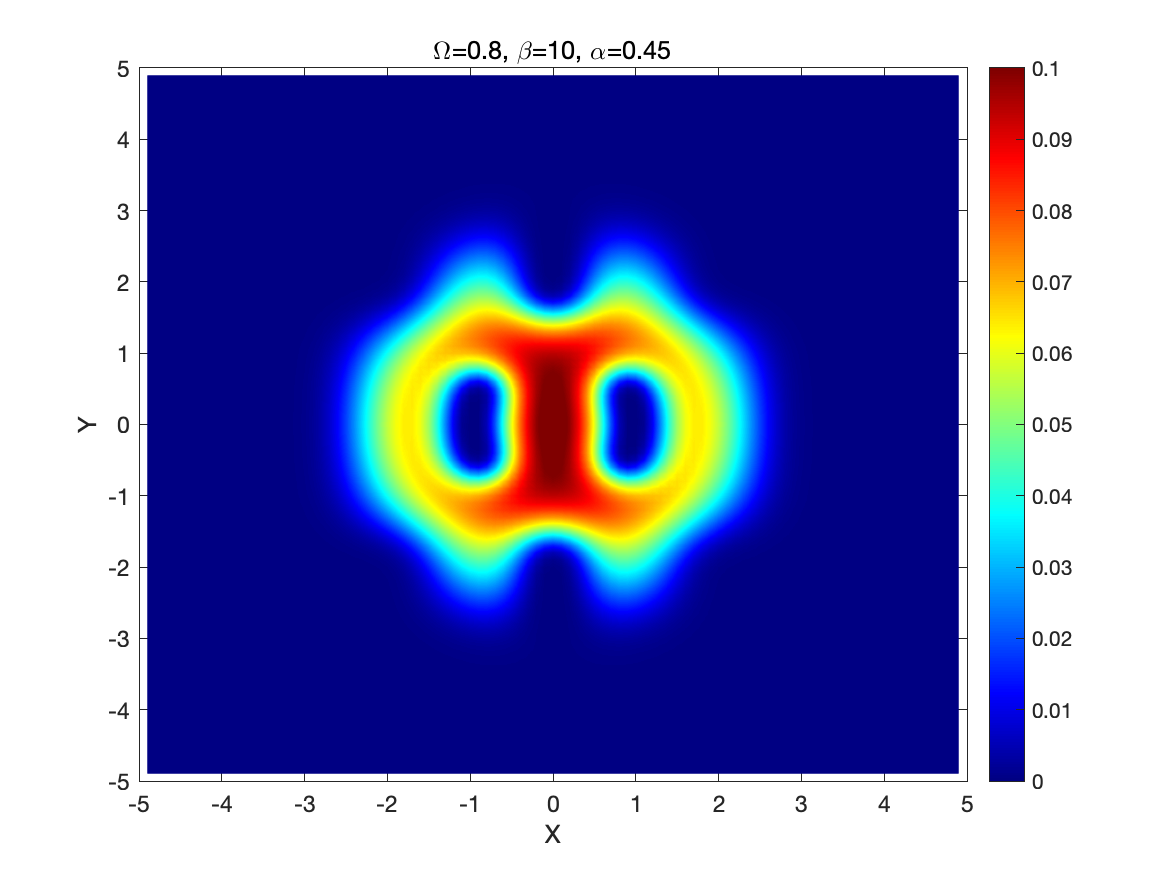}
 }
 \centerline{
  \includegraphics[height=3.8cm,width=4.2cm]{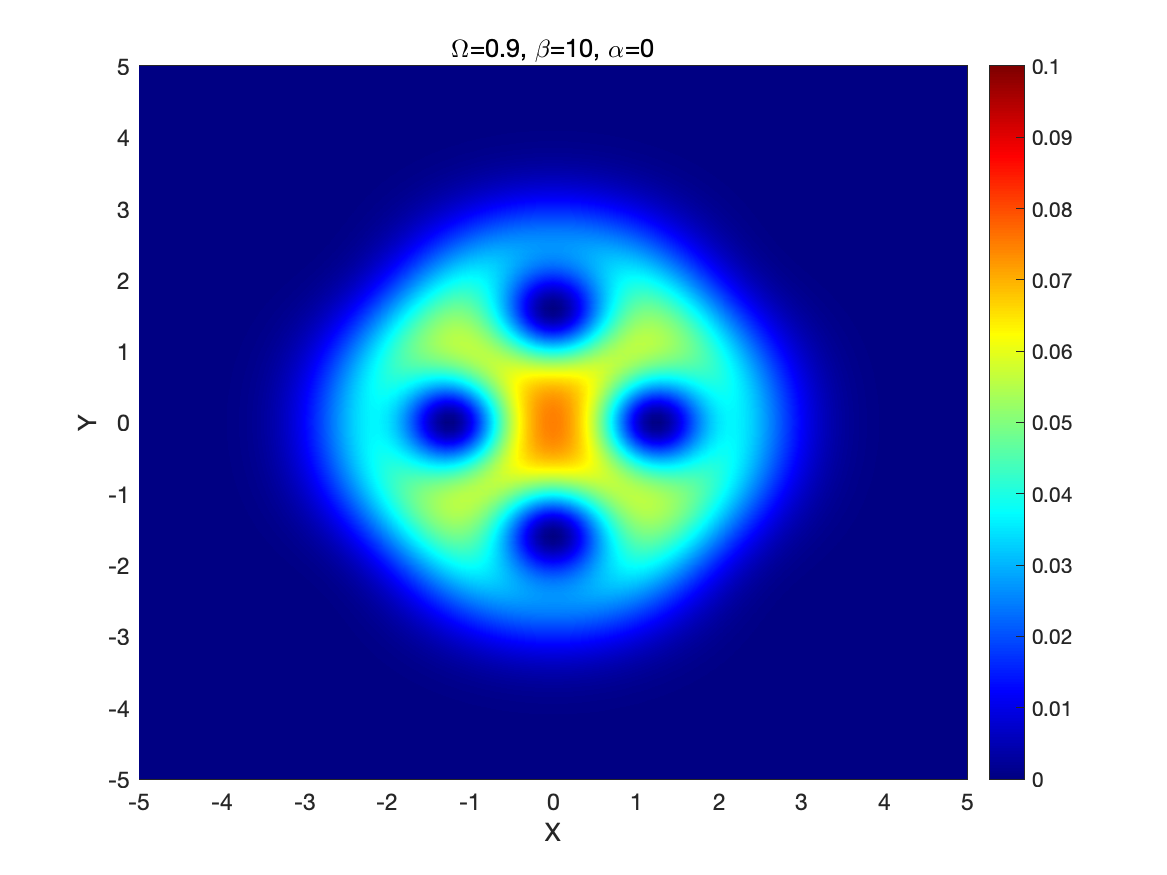}
\includegraphics[height=3.8cm,width=4.2cm]{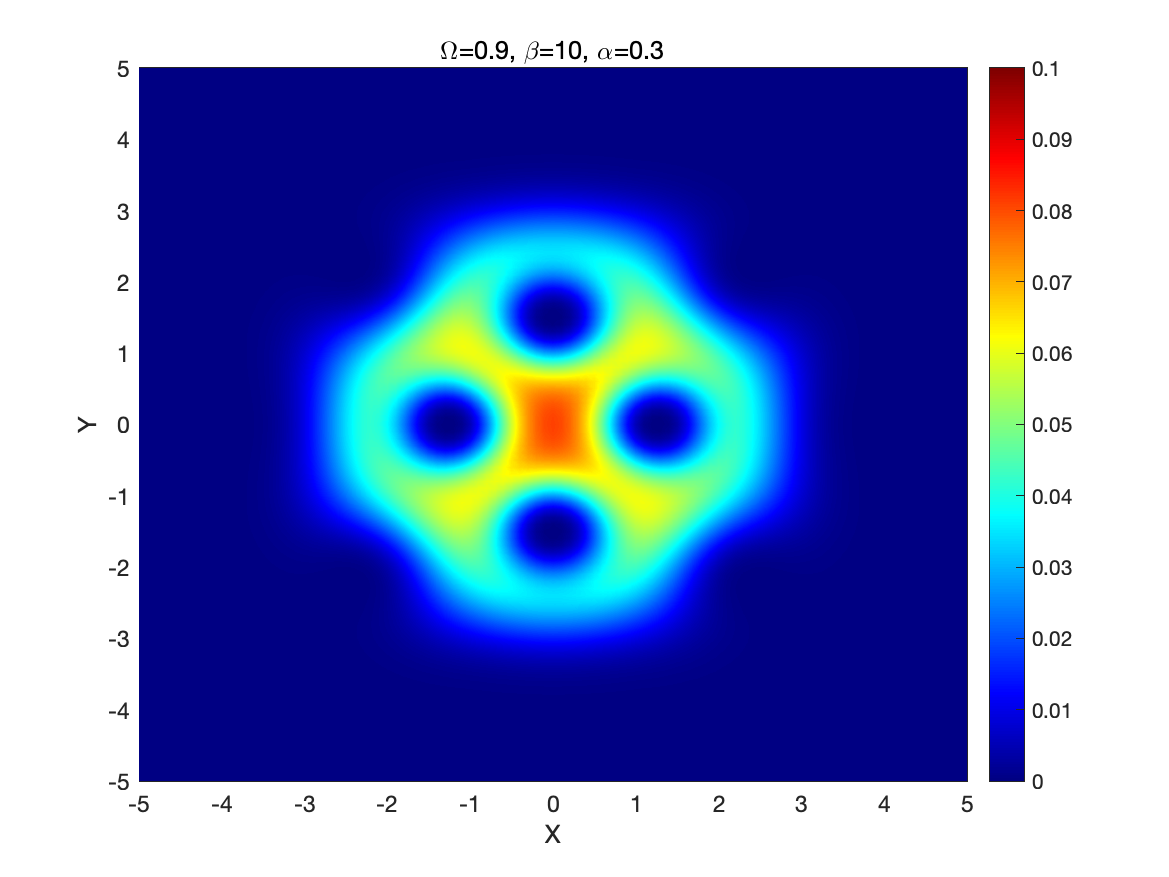}
 \includegraphics[height=3.8cm,width=4.2cm]{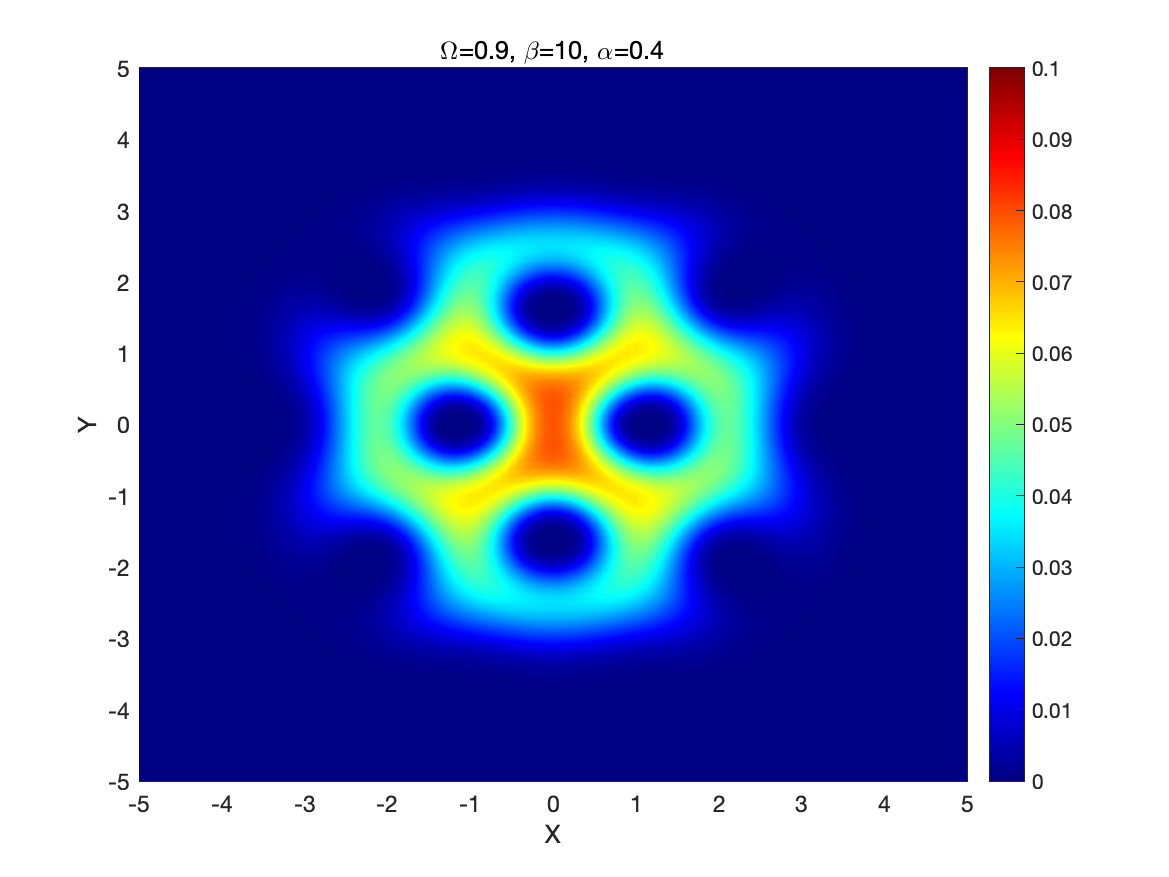}
 \includegraphics[height=3.8cm,width=4.2cm]{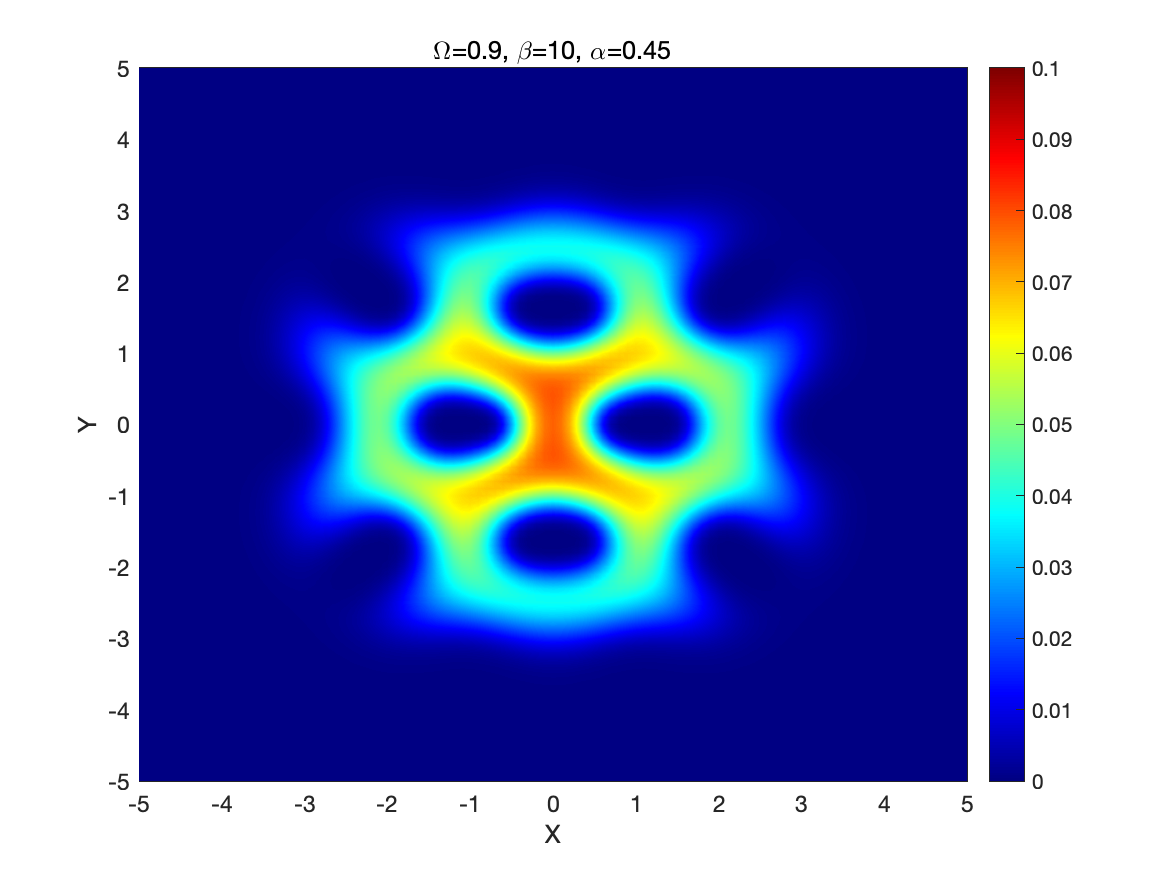}
 }
\caption{Ground states $|\phi_g(x,y)|$ in two-dimensional harmonic potential $V(x,y)  = \fl{1}{2}(x^2+y^2)$,
where $\bt = 10$. From top to bottom:  $\Omega = 0.8$ and $0.9$. 
From left to right:  $\alpha = 0, 0.3, 0.4$ and $0.45$. 
Domain displayed:
$(x,y) \in [-5,5]^2$.}\label{F5}
\end{figure}

\begin{figure}[h!]
\centerline{
\includegraphics[height=3.8cm,width=4.2cm]{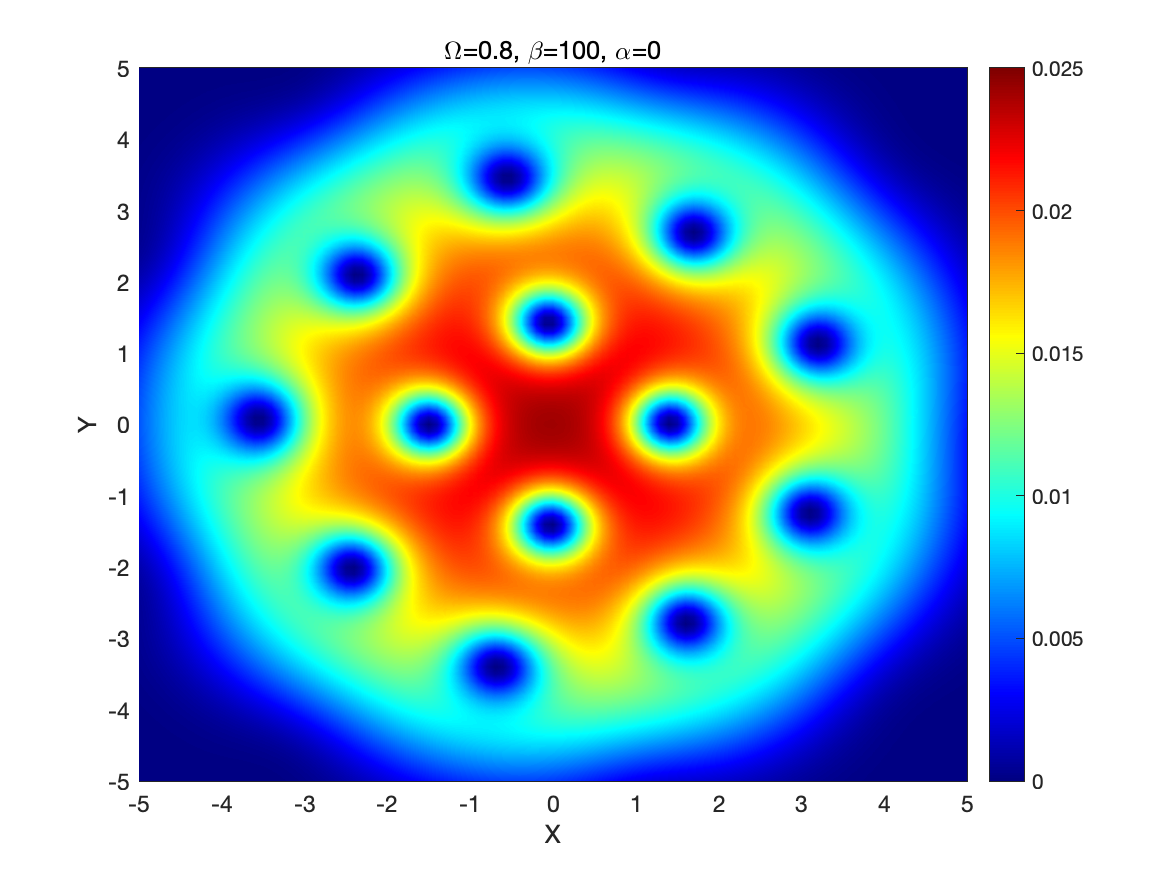}
\includegraphics[height=3.8cm,width=4.2cm]{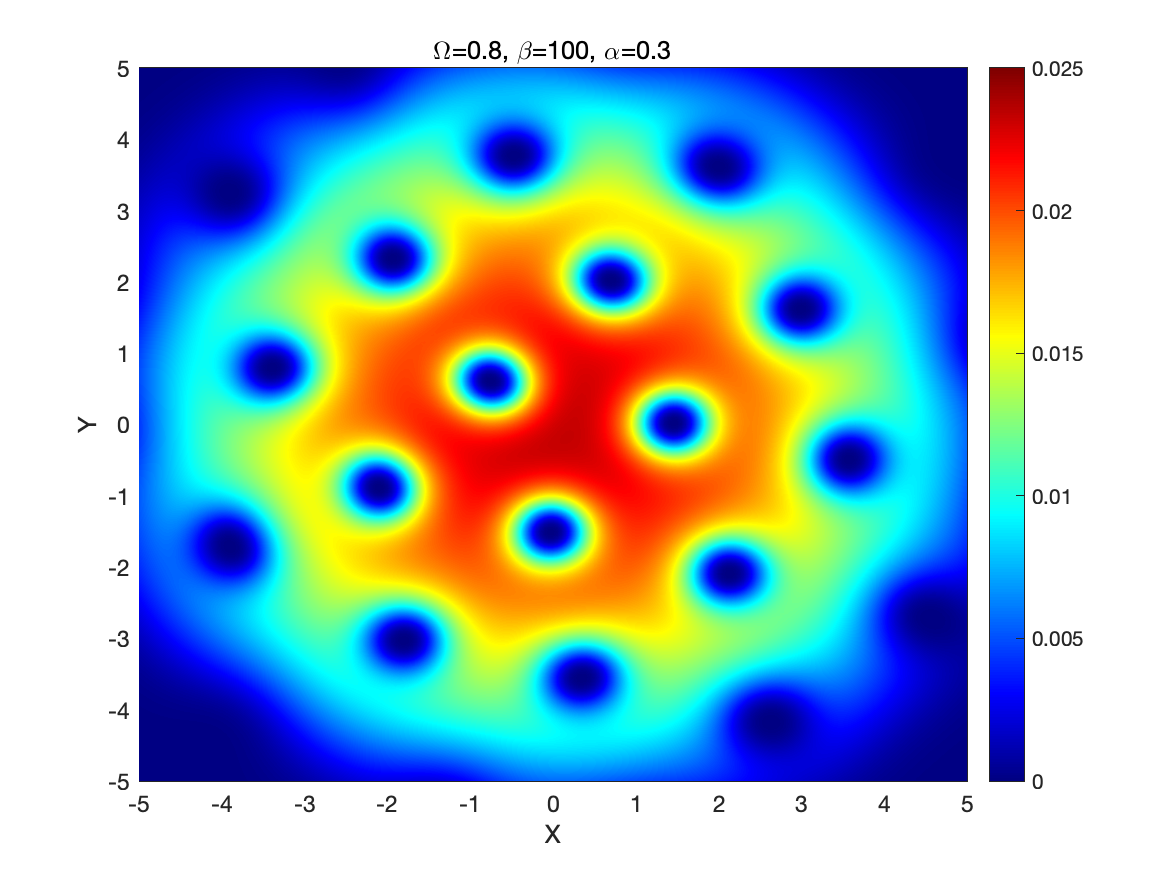}
\includegraphics[height=3.8cm,width=4.2cm]{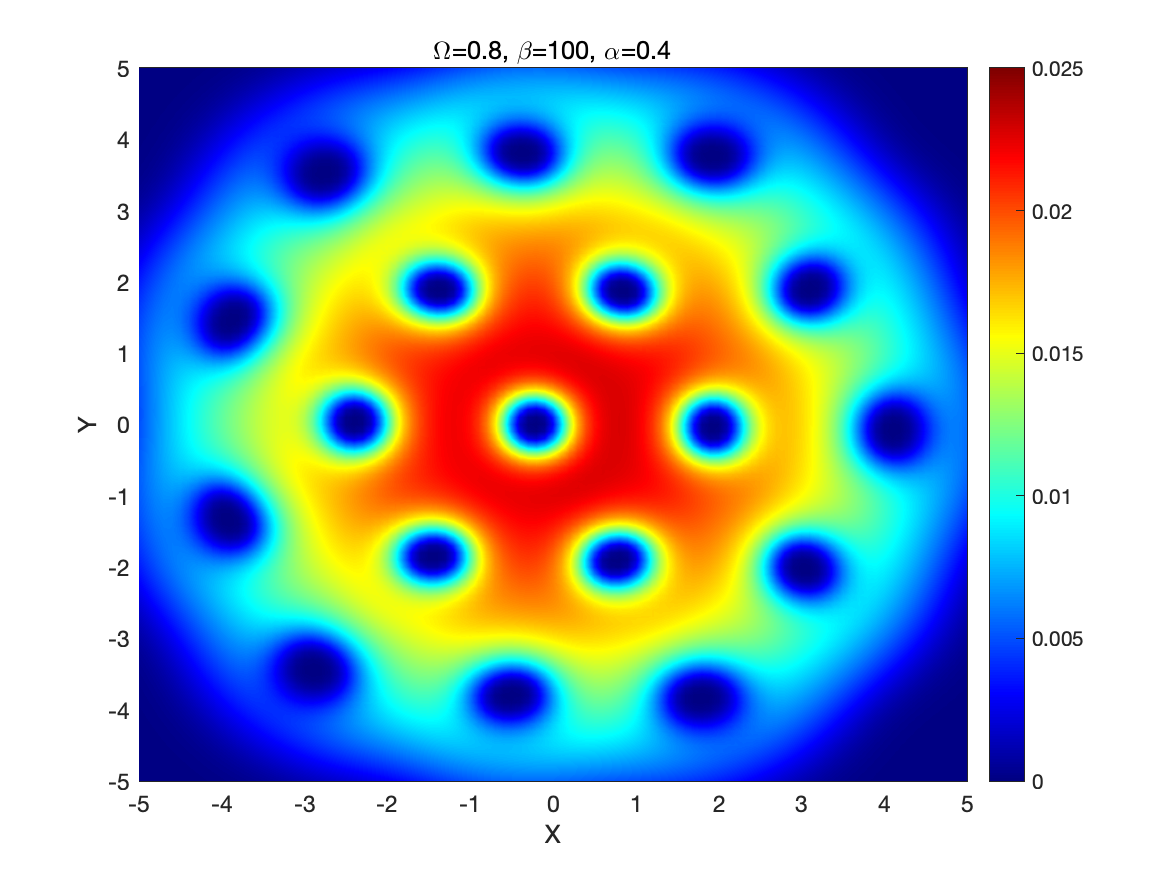}
\includegraphics[height=3.8cm,width=4.2cm]{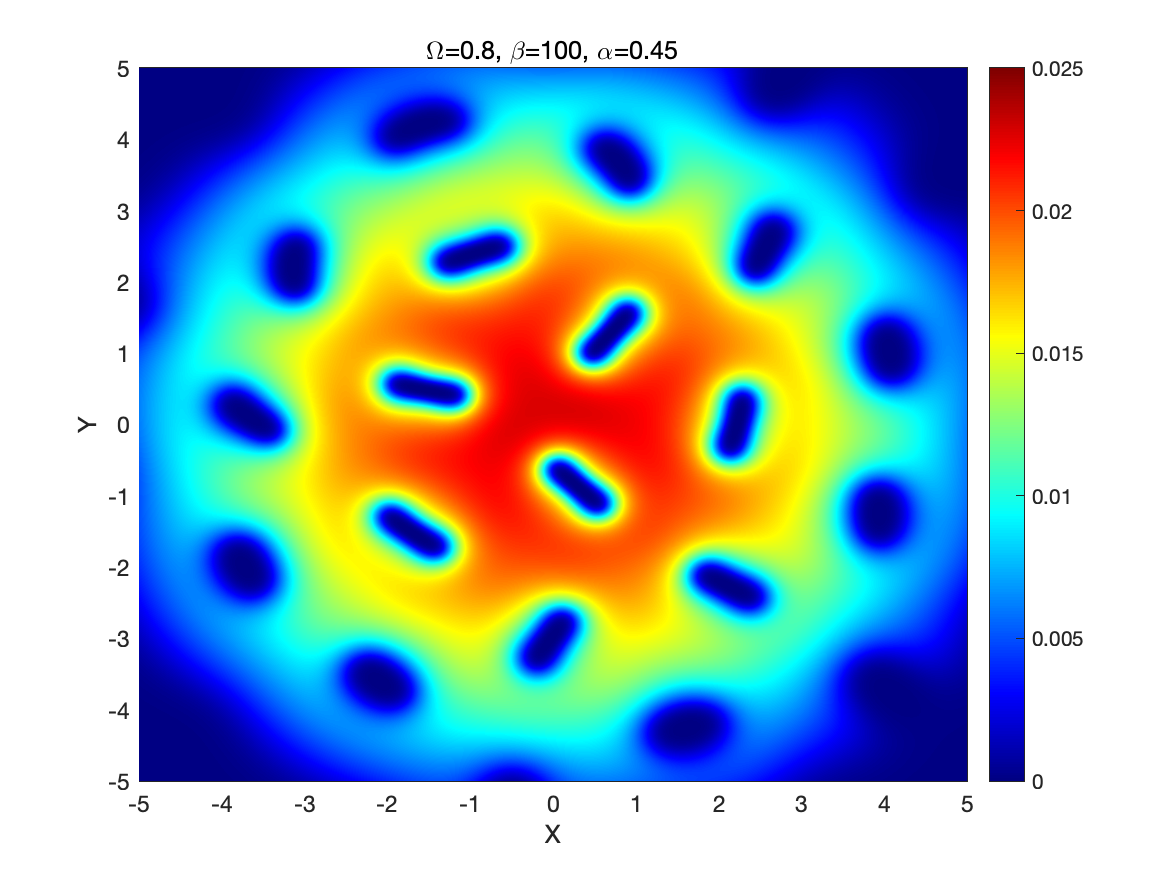}
 }
\caption{Ground states in two-dimensional harmonic potential $V(x,y)  = \fl{1}{2}(x^2+y^2)$,
where $\bt = 100$. From left to right, we choose $\ap = 0$, $0.3$, 0.4 and $0.45$. 
Domain displayed:
$(x,y) \in [-5,5]^2$.}\label{F6}
\end{figure}

\noindent{\bf Example 4 \ } We study the two dimensional ground states under the harmonic potential $V(x,y)=(x^2 + y^2)/2$.   
Figures~\ref{F5} and \ref{F6} show the density plots of the ground states for $\bt = 10$ and  $100$, respectively, where the computational domain is much larger than the region displayed. 
In these plots, dark red corresponds to the highest density, while dark blue indicates regions where the density vanishes. 

From Fig. \ref{F5}, we see that 
the number of vortices increases with the parameter $\Omega$, which is similar to the behavior observed in rotating Bose–Einstein condensates  \cite{Bao2005,Zeng2009}.
However, the presence of the quantum-pressure term in the NLSE \eqref{DNLSE} introduces several new phenomena: 
(i) The quantum pressure term can affect the number of vortices in the ground state. For example,
in Fig. \ref{F5} when $\Og = 0.8$, there is one vortex when $\ap = 0$, but  it increases to two when
$\ap = 0.3$.  (ii) The quantum pressure term can also change the shape of the vortices.  
When $\ap$ is small, all vortices maintain a nearly circular shape. 
However, when $\ap$ is large, all vortices are lengthened, and resolving their structure requires a high–resolution numerical method. 
Fig. \ref{F6} shows  similar results. 
When the parameter $\bt $ is extremely large, $ \Og$ is close to 1 and $\alpha$ is close to $\frac{1}{2}$, the resulting vortex lattice becomes very dense, and highly accurate numerical schemes are needed in this regime.

\bigskip
\section{Summary}
\label{section6}

In this work, we have mathematically and numerically studied the ground states of  non-rotating or rotating unitary Fermi gases.  
Starting from a nonlinear Schr\"{o}dinger equation with a quantum pressure term and an angular momentum rotation term, 
we performed nondimensionalization and dimension reduction to obtain the effective low-dimensional models under suitable parameter regimes of the external trapping potential. 
The existence and uniqueness of ground states were established for both non-rotating and rotating cases. 
To efficiently compute the ground states numerically, we developed a regularized normalized gradient flow method.
The introduced regularization effectively stabilizes the computation, particularly in the presence of both quantum pressure and quantized vortices. 
With this numerical framework, we verified the dimension reduction and Thomas–Fermi approximations proposed in this study.
For general cases, ground states were computed for systems with and without rotation.
In the non-rotating case with a harmonic trapping potential, the numerical results indicate that the influence of the quantum pressure term is significant only when the nonlinear interaction parameter $\beta$ is small.
In the rotating case, we observed that the number of vortices increases with the rotational frequency, a phenomenon consistent with that in rotating Bose–Einstein condensates \cite{Zeng2009}.
However, the inclusion of the quantum pressure term leads to distinct vortex structures and notable modifications in vortex lattice configurations, offering new physical insights into the behavior of quantized vortices in unitary Fermi gases. 
Future work will focus on investigating the dynamics of the nonlinear Schrödinger equation for unitary Fermi gases and exploring the evolution of vortex patterns.

\bigskip
\noindent{\large\bf Acknowledgements}  \ We thank helpful discussions with Prof. Weizhu Bao in
the subject.  Y. Cai acknowledges support  from the National Natural Science Foundation of China under grants 12571412, 12171041.
X. Ruan acknowledges support  from the National Natural Science Foundation of China under grant  12201436.
Y. Zhang acknowledges partial support from the U.S. National Science Foundation under grant DMS–1913293.


\end{document}